\documentclass[sigconf]{acmart}
\acmConference[ESEC/FSE 2020]{The 28th ACM Joint European Software Engineering Conference and Symposium on the Foundations of Software Engineering}{8 - 13 November, 2020}{Sacramento, California, United States}

\AtBeginDocument{%
  \providecommand\BibTeX{{%
    \normalfont B\kern-0.5em{\scshape i\kern-0.25em b}\kern-0.8em\TeX}}}

\usepackage{amsmath}

\usepackage{amssymb}
\usepackage{amsfonts}
\usepackage{hyperref}
\usepackage{algorithmic} 
\usepackage{graphicx}
\usepackage{textcomp}
\usepackage{xcolor}
\usepackage{xspace}
\usepackage[ruled,linesnumbered]{algorithm2e}
\usepackage{multirow}
\usepackage{booktabs}
\usepackage{soul}
\usepackage{balance}
\usepackage{tcolorbox}
\renewcommand\footnotetextcopyrightpermission[1]{}

\newcommand\tool{\textsc{DeepJanus}\xspace}

\newboolean{showcomments}
\setboolean{showcomments}{true}         
\ifthenelse{\boolean{showcomments}}
  {\newcommand{\nb}[2]{
  \fbox{\bfseries\sffamily\scriptsize#1}
     {\sf\small$\blacktriangleright$\textit{\textcolor{red}{#2}}$\blacktriangleleft$}
   }
  }
  {\newcommand{\nb}[2]{}
   
  }

\begin{document}

\title{Model-based Exploration of the Frontier of  Behaviours for Deep Learning System Testing}

\author{Vincenzo Riccio}
\affiliation{%
  \institution{Software Institute - USI}
  \city{Lugano}
  \country{Switzerland}}
 \email{vincenzo.riccio@usi.ch}

\author{Paolo Tonella}
\affiliation{%
  \institution{Software Institute - USI}
  \city{Lugano}
  \country{Switzerland}}
 \email{paolo.tonella@usi.ch}
  
\begin{abstract}
With the increasing adoption of Deep Learning (DL) for critical tasks, such as autonomous driving, the evaluation of the quality of systems that rely on DL has become crucial. Once trained, DL systems produce an output for any arbitrary numeric vector provided as input, regardless of whether it is within or outside the validity domain of the system under test. Hence, the quality of such systems is determined by the intersection between their validity domain and the regions where their outputs exhibit a misbehaviour. 

In this paper, we introduce the notion of frontier of behaviours, i.e., the inputs at which the DL system starts to misbehave. If the frontier of misbehaviours is outside the validity domain of the system, the quality check is passed. Otherwise, the inputs at the intersection represent quality deficiencies of the system. We developed \tool, a search-based tool that generates frontier inputs for DL systems. The experimental results obtained for the lane keeping component of a self-driving car show that the frontier of a well trained system contains almost exclusively unrealistic roads that violate the best practices of civil engineering, while the frontier of a poorly trained one includes many valid inputs that point to serious deficiencies of the system.
\end{abstract}



\maketitle

\section{Introduction}
\label{introduction}

Deep Neural Networks (DNNs) have been applied successfully to complex tasks such as image processing, speech recognition and natural language analysis. The range of applications of DNNs is huge and includes autonomous driving, medical diagnosis, financial trading and automated customer services. As a consequence, the problem of testing Deep Learning (DL) systems to ensure their dependability has become critical. 

Existing approaches to generate tests for DL systems can be split into two groups: (1) techniques that directly manipulate the raw input data~\cite{PeiCYJ17,ZhangZZ0K18,WickerHK18} and (2) techniques that derive input data from a model of the input domain~\cite{AbdessalemNBS16,GambiMF19}. In the former case the result is a so called \textit{adversarial example}, while in the latter case it can be regarded as a \textit{critical test scenario}. In both cases, test generation is guided by the criticality of the produced inputs, measured either directly as a misclassification/inconsistent behaviour~\cite{AbdessalemNBS16,GambiMF19,WickerHK18} or mediated by a proxy such as surprise~\cite{KimFY19} or neuron coverage~\cite{Ma-ASE-2018,PeiCYJ17}.
Adversarial examples, e.g. images obtained by manipulation of the pixels of an image taken from the camera of an autonomous car, may produce very unlikely (or even impossible) cases, whose resolution might have no impact on the system's reliability. Critical test scenarios obtained by model-based input generation tend to be more realistic. However,  the existing model-based approaches  do not aim at covering thoroughly and characterising the  region where the DL system misbehaves.

The ISO/PAS standard 21448~\cite{pas21448} on safety of autonomous and assisted driving prescribes that unsafe situations should be identified and be demonstrated to be sufficiently implausible. When unsafe situations are plausible, countermeasures must be adopted. Manual identification of unsafe conditions for DNNs is challenging, because their behaviour cannot be decomposed via logical conditions, as done e.g. with root cause analysis~\cite{iso26262}. This motivates our work on the automated identification of the frontier of behaviours of DL systems: we aim to support engineers in identifying and checking the plausibility of the frontier of behaviours.

In this paper, we introduce a novel way to assess the quality of DL systems, based on a new notion: the \textit{frontier of behaviours}. The frontier of behaviours of a DL system is a set of  pairs of  inputs that are similar to each other and that trigger different behaviours of the DL system. It represents the border of the input region where the DL system behaves as expected. For instance, the frontier of a classifier of hand-written digits consists of the pairs of similar digits that are classified differently (one correctly and the other incorrectly). The frontier of behaviours of a low quality DL system may include pairs of inputs that intersect the validity domain, being similar to nominal cases for which the system is expected to behave correctly according to the requirements. 
On the contrary, a DL system of high quality will start to misbehave on inputs that deviate substantially from the nominal ones, with small or no intersection with the validity domain (e.g. a digit ``5'' is misclassified only when it becomes unrecognisable or indistinguishable from another digit, such as ``6'').

We have adopted a model-based input generation approach to produce realistic inputs, under the assumption that a high fidelity model of the input is available for the DL system under test. There are several domains in which the development of input models is standard practice, among which safety-critical domains such as automotive and aerospace engineering~\cite{Larman1997}. In other domains, such as image classification, input models can be constructed (e.g. in Unity~\cite{unitygameengine}) or reverse engineered. Our tool \tool implements a multi-objective evolutionary algorithm to manipulate the input model,  with the overall goal of achieving thorough exploration of the frontier of behaviours. To this aim, one of its two fitness functions  promotes diversity, so as to spread the solutions along the entire frontier, and minimises the distance between the elements in each pair. The other fitness function pushes the solutions to the frontier. The output of \tool provides developers with a thorough and human-interpretable picture of the system's quality. In fact, the elements of each pair in the frontier may be deemed as within or outside the validity domain of the system (in the latter case, they are irrelevant for the reliability of the DL system). When used to compare alternative DL systems that solve the same problem, metrics of the frontier size (e.g. its radius) are useful to show quantitatively if the region contained in one frontier is substantially smaller/larger than the region contained in the other.

The proposed technique was evaluated on both a classification problem (hand-written digit recognition) and a regression problem (steering angle prediction in a self-driving car). The frontier of the digit classifier was evaluated by $20$ human assessors recruited on a crowdsourcing platform. Results show that a high quality classifier has a smaller intersection with the validity domain with respect to a poorly trained one. The frontier of the self-driving car was evaluated by assessing the conformance of the shapes of the roads at the frontier to the guidelines for the design of American highways~\cite{aashto2018}. Frontier roads obtained for a high quality system violate such guidelines, showing that the system misbehaves only in extreme cases. Such results were confirmed by quantitative measures of the frontier radius, which was  larger for the high quality than for the low quality DL system, and qualitative assessment of the frontier images/roads, which are more challenging for humans when taken from the high quality system frontier. 

We compared our results with those produced by DLFuzz~\cite{GuoJZCS18}, a tool that generates boundary adversarial inputs by pixel manipulation, and found that it generates corner cases that are more concentrated and less realistic than those of \tool. 

\section{Background}
\label{background}
\subsection{Deep Learning Systems}

In this work, we refer to software systems that include one or more DNNs as DL systems \cite{Ma-ASE-2018}. Their behaviour is defined both by the code that implements them and by the data used to train their DNN components.

A DNN can be considered as a black-box component that transforms a numeric input vector into a numeric output. It can accomplish various tasks, such as the prediction of the steering angle of a self-driving car starting from the image captured by a camera sensor \cite{BojarskiNVIDIA16}. In a regression problem the output is a continuous value, whereas in a classification problem the output is a discrete class.

A DNN consists of a collection of computation units, called \textit{neurons}, organised into \textit{layers} that are connected sequentially (i.e. neurons of layer $n$ are only connected to neurons of layer $n+1$). Each connection of the network has a \textit{weight}, which determines the propagation of a neuron's output to the next neuron. Among the layers of a DNN, the input layer receives external data, the output layer  produces the final result, while internal, hidden layers perform intermediate processing (e.g. feature extraction). Each neuron computes its output by applying an activation function (e.g., sigmoid) to the weighted sum of its inputs. 

To accomplish a task, DNNs are iteratively trained through a large set of labelled training data. During  training, a DNN learns how to perform a prediction, i.e. a label for classification problems or a real value for regression problems, by adjusting the weights of the network. The number of training iterations in which the whole training set is processed by the network is a hyper-parameter called \textit{epochs}. The number of epochs influences how the network fits  the training data and how it will be able to generalise to unseen inputs.
Another fundamental  hyper-parameter is the \textit{learning rate}, which defines the amount of  corrections that are applied to the weights at each training iteration. 

\subsection{Evolutionary Search and Novelty Search}

Evolutionary algorithms are a family of meta-heuristic optimisation algorithms that evolve a population of \textit{individuals} (i.e. candidate solutions to an optimisation problem) by means of genetic operators such as \textit{mutation} and \textit{crossover}. A \textit{fitness function} provides an approximate, heuristic distance of each candidate solution from the searched optimum. During evolution, the best individuals  are selected for the next population based on the fitness function values. 

Multi and many objective evolutionary algorithms generalise the basic evolutionary algorithms to multiple fitness functions. Since, in such a case, there is no single dimension on which to compare individuals during selection, the best ones are obtained by Pareto front analysis as those that are not dominated by any other individual. Multi and many objective genetic algorithms have proved to be particularly effective in test case generation~\cite{PanichellaKT18, MaoHJ16}.

The solutions found by search algorithms might be concentrated in a small portion of the input space, especially when the search landscape includes local optima with a large basin of attraction. If the goal is not only to find good solutions according to the fitness functions, but also to find solutions spread across the entire input space (as in our case), evolutionary algorithms can be combined with \textit{novelty search}. Novelty search algorithms  reward individuals that exhibit diversity of  behaviours, instead of promoting only those that contribute to progress toward the optimum~\cite{LehmanS11, MarculescuFT2016}. They trade off a lower pressure toward optimal fitness values with a higher diversity in the population being evolved.

\section{Motivating Example}
\label{motivating}

\begin{figure}
  \includegraphics[width=\linewidth]{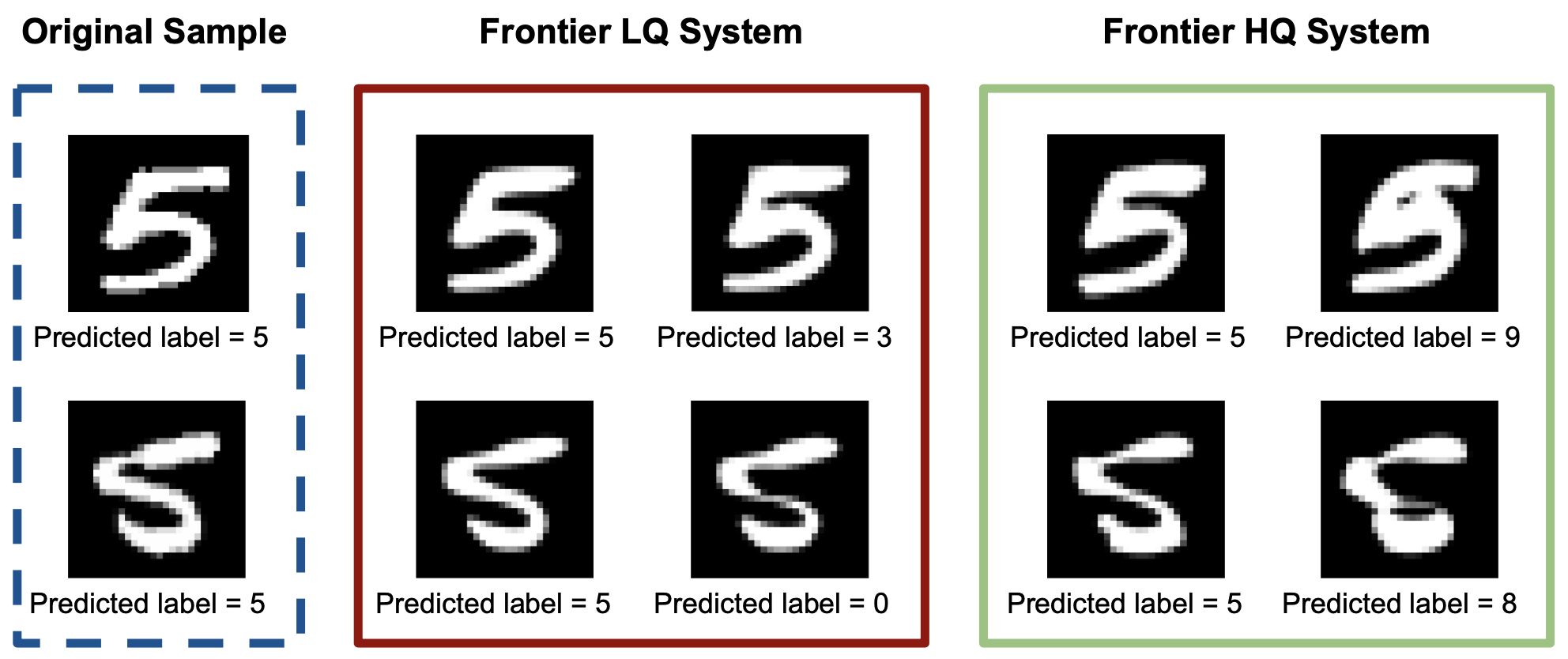}
  \caption{Four pairs of elements in the frontiers of the considered DL systems. The first column shows two original samples from the MNIST data set. The second and third columns show the pairs in the frontier of system LQ; fourth and fifth columns show the frontier pairs of system HQ.}
  \label{fig:comparison}
\end{figure}

In this Section, we provide a motivating example that shows how the frontiers of behaviours can help characterising the quality of DL systems. Let us consider the frontiers of behaviours of two DL systems that perform the same task but exhibit different levels of quality in terms of test accuracy (namely HQ: High Quality: LQ: Low Quality). Both systems consist of a classifier of handwritten digits that predicts which digit is represented by an input image. In a classification problem such as this one, the frontier is represented by pairs of similar inputs that are classified differently (one correctly, the other incorrectly).

To assess the difference between the frontiers of these two  systems, we consider two images of handwritten digits taken from the  MNIST~\cite{LecunBBH98} dataset that are labeled correctly (i.e. as number ``5'') by both systems. 
They are shown in the first column of Fig. \ref{fig:comparison}.
Then, we apply slight changes to the shape of the two inputs. Technically, this is achieved by first extracting a vector model of the digits and then manipulating the control points of such model. The result consists of two pairs of samples in the frontier of each considered system, i.e. LQ (second and third column) and HQ (fourth and fifth column). 

We can notice that the inputs in the frontier of LQ are very similar to the original samples. Moreover, all the misclassified inputs in its frontier are still clearly recognisable as digit ``5''. Instead, the frontier of HQ contains inputs that are probably challenging to classify even for humans. In particular, the first element of the fifth column has the general shape of a five, but it could also be considered  as a nine, since the upper part of the figure forms a circle. The second element of the fifth column does not look like any reasonably classifiable digit, despite its similarity with the corresponding member of the pair on the other side of the frontier.

To summarise, the frontier of a low quality DL system is expected to contain samples that are quite close to those that the system is supposed to classify correctly, indicating a poor generalisation capability. Differently, the frontier of a high quality DL system includes cases that are difficult or impossible to handle even for humans, being outside the validity domain.

\section{Model-based Input Representation}
\label{model}

We aim at generating inputs at the behavioural frontier of a DL system and we want them to be realistic and representative.
Therefore, we adopt a model-based approach that produces test inputs starting from a model representation of the input domain and enforces the compliance with domain-specific constraints.
This may require the transformation of a concrete input into an abstract model that can be manipulated by the exploration algorithm, in case no domain specific model of the input is available. The transformation from models to concrete inputs is instead always required.

To illustrate how our approach works in practice,
we consider both an exemplary classification problem and a regression problem. 
The classification problem consists of handwritten digit recognition,
while the regression problem is steering angle prediction for self-driving cars. In the latter case, we focus on systems that perform behavioural cloning, i.e. the DL component learns the lane keeping behaviour from a human driver~\cite{BojarskiNVIDIA16}. In detail, the DL system is able to autonomously keep the lane since it contains a DNN that is trained with images captured by the camera sensors of the car, paired with the steering angles provided by a driver.

\subsection{Image Classification} 
We use the inputs available from the MNIST database~\cite{LecunBBH98} and originally encoded as 28 x 28  images~\cite{LecunBBH98}, with greyscale levels that range from 0 to 255.
We adopt Scalable Vector Graphics (SVG)\footnote{\url{https://www.w3.org/Graphics/SVG/}} as their model representation. SVG is an XML-based vector image format for two-dimensional graphics that can represent shapes as the combination of cubic and quadratic B\'ezier curves~\cite{Hazewinkel97}. By modelling handwritten digits as a combination of B\'ezier curves, we ensure that the smoothness and curvature of handwritten shapes is preserved and that images remain realistic even after (minor) manipulation of the B\'ezier curve parameters.

\begin{figure}
\centering
  \includegraphics[width=\linewidth]{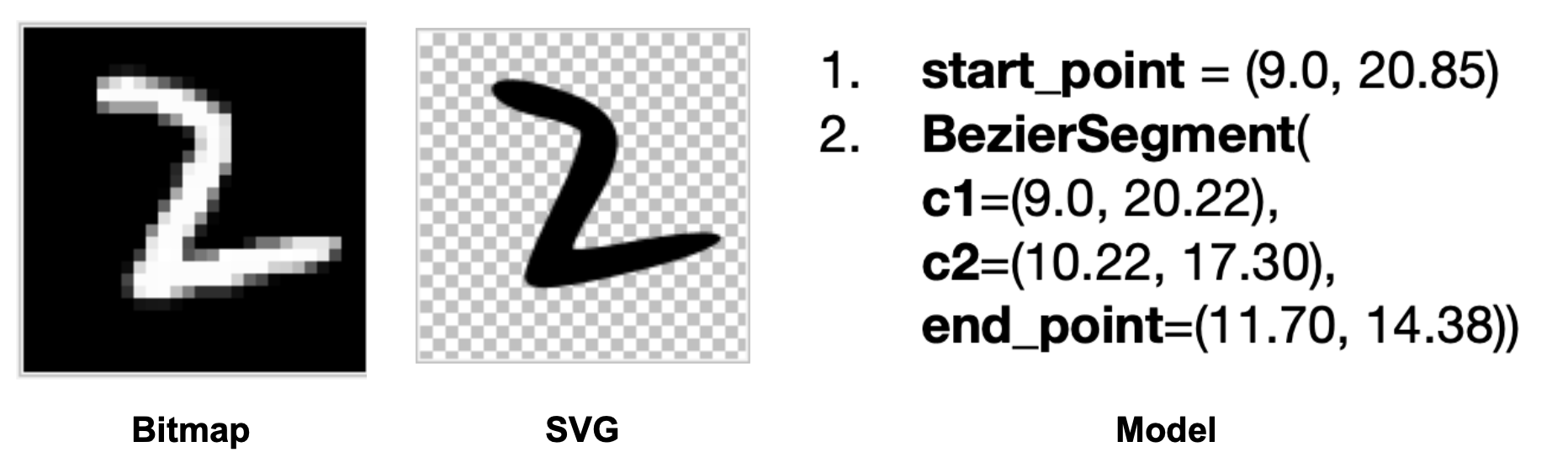}
  \caption{Bitmap and vector image;  model representation of the image returned by Potrace}
  \label{fig:svg_model}
\end{figure}

To transform an original input image into its SVG model representation, we use the Potrace algorithm~\cite{Selinger03}. This algorithm performs a sequence of operations, including binarisation, despeckling and smoothing, to produce a smooth vector image starting from a bitmap. Figure~\ref{fig:svg_model} shows an MNIST image paired with its SVG model and its description. The control parameters that determine the shape of the modelled digit are: the start point, the end point and the control points \textit{c1} and \textit{c2} that define each B\'ezier segment.

In the other direction, we use rasterisation to transform a vector model into a 28 x 28  grayscale image. This operation exploits the functionality offered by LibRsvg\footnote{\url{https://wiki.gnome.org/Projects/LibRsvg}} and Cairo\footnote{\url{https://www.cairographics.org}}, two popular open source graphic libraries.

\subsection{Steering Angle Prediction} \label{sec:roadmodel}

We consider a self-driving car that is trained and tested in the BeamNG~\cite{beamng_research} simulation environment. It features an accurate driving physics engine and it is freely available and research-oriented.

The input to the steering angle predictor is an image captured by the onboard sensor camera in the simulated environment. 
Therefore, the test input is determined by the scenario in which the car drives.
Such simulated scenario can be modelled as the composition of the roads, the driving task (i.e., start point, end point and lane to keep), and the environment, which includes the weather and lightness conditions. 

For the sake of simplicity, let us consider scenarios consisting of single plain asphalt roads surrounded by green grass on which the car has to drive keeping the right lane. The environment is always set to a clear day without fog. The roads are composed of two lanes with fixed width in which there is a yellow center line plus two white lines that separate each lane from the non-drivable area.

\begin{figure}
\centering
  \includegraphics[width=\linewidth]{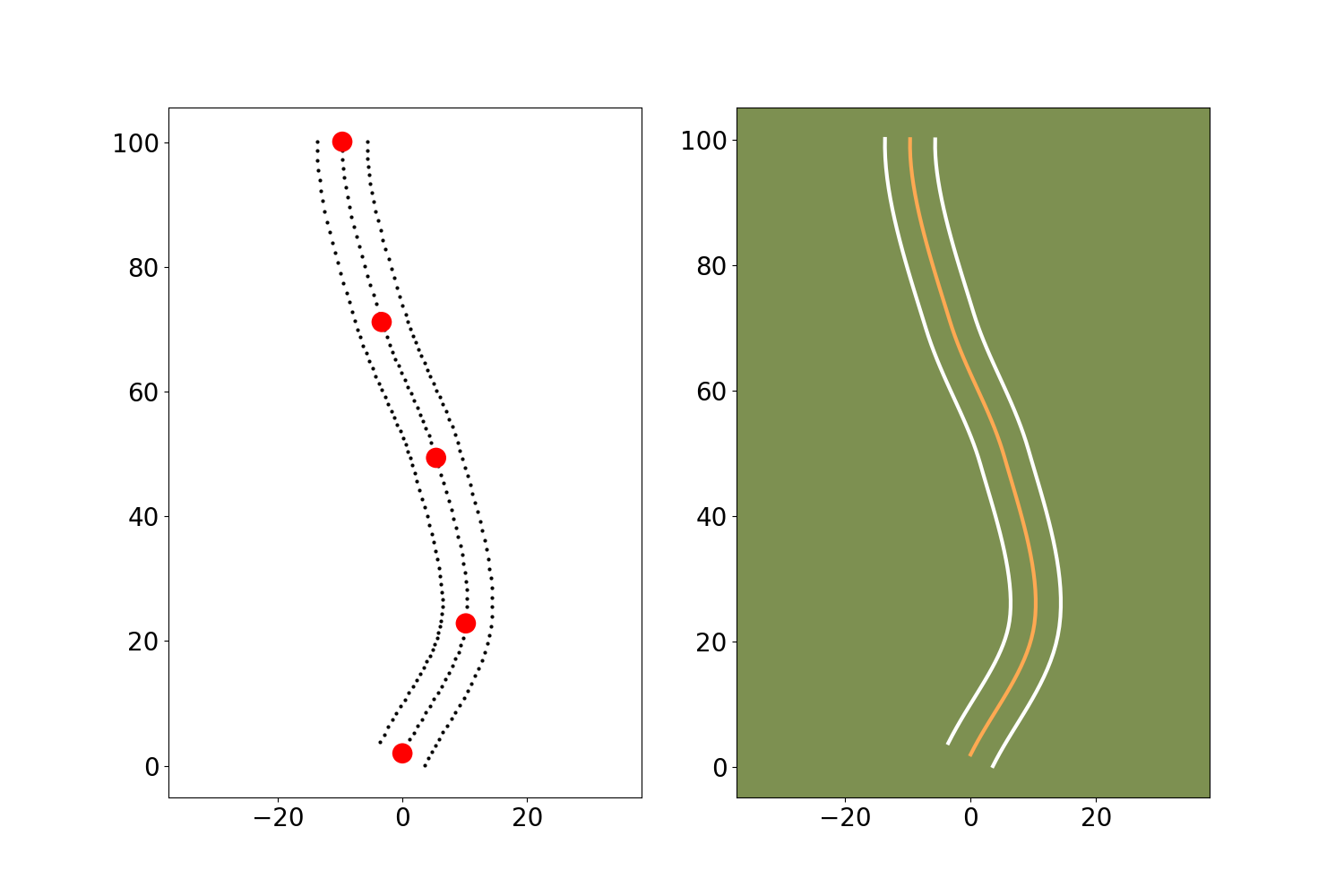}
  \caption{The model of a road and the corresponding road}
  \label{fig:road_model}
\end{figure}

Abstractedly, a road can be represented as a sequence of contiguous points in a bi-dimensional space (assuming constant elevation). 
To produce a smooth and realistic shape for the road being modelled, we use Catmull-Rom cubic splines~\cite{CatmullRom74}  and then we interpolate such curves to obtain the 2D point sequence. \autoref{fig:road_model} shows the splines that define a road as well as its interpolated 2D points (marked as grey dots). The control parameters that determine the shape of the splines in \autoref{fig:road_model} are the coordinates of the control points of the center line spline (marked as larger red  dots).

\begin{figure}
\centering
  \includegraphics[width=0.8\linewidth]{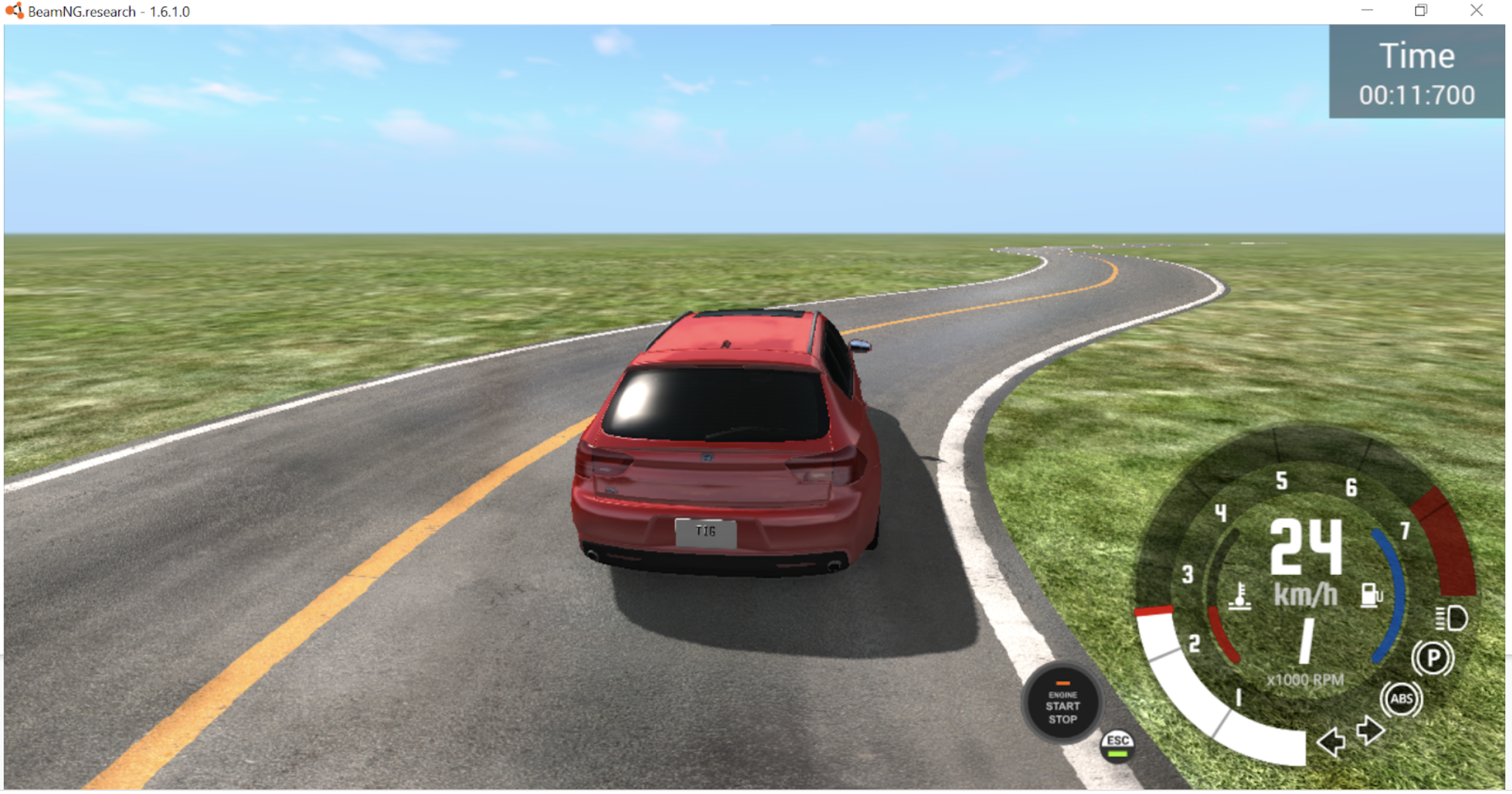}
  \caption{A test case rendered by the BeamNG simulation engine is composed by the road, the driving task, the environment and the car}
  \label{fig:simulation}
\end{figure}

The concrete representation of the driving scenario is strictly dependent on the simulator. 
BeamNG exposes an intuitive API for programmatically configuring virtual roads and controlling the simulations\footnote{\url{https://github.com/BeamNG/BeamNGpy}}. In BeamNG, a scenario is described by a JSON file that contains the set of points to render the roads. The simulation engine renders the road by creating polygons starting from the points provided in the scenario description and sets up the environment, as shown in \autoref{fig:simulation}.

To transform the abstract model into the road to be rendered in the simulator, we calculate its points by exploiting the recursive algorithm for the evaluation of Catmull-Rom cubic splines proposed by Barry and Goldman~\cite{BarryG88} and the functionality offered by the Shapely library for manipulation and analysis of planar geometric objects\footnote{\url{https://github.com/Toblerity/Shapely}}.
We also enforce the following domain specific constraints: (1) the start point and the end point of a driving task should be different, (2) the road should fall within a square bounding box of fixed size, and (3) a road should not self-intersect.

\section{The DeepJanus Technique}
\label{technique}

\tool explores the behavioural space of a DL system to find pairs of inputs at its frontier: one input on which the DL system behaves as expected, and another similar input on which it misbehaves. By generating a pair of similar inputs that trigger different behaviours, we ensure that the failure-inducing inputs are close to the validity domain and are likely to represent valid corner cases on which the system misbehaves. Otherwise, by generating single inputs that trigger misbehaviours, without staying close to corresponding ones for which the system behaves well, it would have been more likely to produce uninteresting test cases that are far from the frontier and do not intersect the validity domain.

\tool aims at exploring the frontier at large, i.e., as thoroughly as possible, so as to report a broad picture of the boundary behaviours to developers. To perform such  exploration, it aims at producing inputs at the frontier of  behaviours and at maximising the diversity among the elements that are moved toward the frontier, so as to achieve thorough frontier exploration. At the same time, it also maintains high similarity within each pair of inputs crossing the frontier. Therefore, the problem solved by \tool can be cast as a multi-objective search problem~\cite{HarmanMAZ12}. To obtain a diverse set of solutions, we hybridise traditional multi-objective search-based algorithms~\cite{DebAM02} with novelty search~\cite{Mouret09}. The idea is to measure the diversity between the population being evolved and the archive of the best individuals. 

Algorithm \ref{algo} outlines the top level steps implemented in \tool. Our algorithm is based on  NSGA-II~\cite{DebAM02}, a multi-objective evolutionary search algorithm quite popular in search-based software testing research~\cite{PanichellaKT18,YooH07,YooH10,MaoHJ16,LakhotiaHM07}, extended with:
(1) hybridisation with novelty search, achieved by defining a fitness function that includes a measure of sparseness of the solutions (see Section \ref{sec:ff1}); (2) use of an archive, to avoid cycling and to promote frontier exploration at large (lines 5 and 15 of \autoref{algo}); (3) use of re-population, to escape from stagnation (line 13 of \autoref{algo}). Moreover, we defined domain specific mutation operators to evolve the candidate solutions.

We implemented \tool in Python on top of the DEAP evolutionary computation framework (v. 1.3.0)~\cite{FortinDGPG12}. The code of \tool is available online as open source~\cite{tool}.

\begin{algorithm}[t]
\begin{small}
\caption{Overall algorithm of \tool}
\label{algo}

\SetKwInOut{Input}{Input}
\SetKwInOut{Output}{Output}
\Input{$S$: set of input seeds \\ $g_{max}$: max number of generations \\ $popsize$: population size}
\Output{$A$: archive of best individuals at the frontier}

\textit{generation g} $\gets$ 0\;

\textit{A} $\gets$ $\emptyset$\;

\textit{population P} $\gets$ \textsc{InitialisePopulation}($S$, $popsize$)\;
\textsc{Evaluate}(\textit{P})\;
\textit{A} $\gets$ \textsc{UpdateArchive}(\textit{P})\;
\tcc{assign  crowding distance to  individuals}
\textit{P} $\gets$ \textsc{Select}($P$, $popsize$)\;

\While{$g$ $<$ $g_{max}$ } {
$g$ $\gets$ $g + 1$\;
\tcc{Tournament selection based on dominance and crowding distance}
\textit{offspring Q} $\gets$ \textsc{SelTourDCD}($P$, $popsize$) \;
\ForEach{$q$ $\in$ Q}{
     $q$ $\gets$ \textsc{Mutate}($q$) \;
}

\tcp{substitute the most dominated individuals}
\textit{P} $\gets$ \textsc{Repopulation}($P$, $S$, $A$)\;
\textsc{Evaluate}($P \cup Q$)\;
\textit{A} $\gets$ \textsc{UpdateArchive}($P \cup Q$)\;
\textit{P} $\gets$ \textsc{Select}($P \cup Q$, $popsize$)\;
}
\Return(\textit{A})
\end{small}
\end{algorithm}

\subsection{Fitness Functions}

The algorithm optimises two fitness functions, which measure respectively the quality of an individual (consisting of its  cross-pair diversity and within pair similarity) and the closeness of the inputs to the frontier of behaviours.

\subsubsection{Quality of an individual} \label{sec:ff1}
The quality of an individual is measured by two factors, namely (1) the distance between the two members of a pair and (2) the sparseness of an individual with respect to the individuals in the archive measure the quality of a given individual. Since both are distances between inputs, we can combine these two measures additively into the fitness function $f_1$, to be maximised: 

\begin{equation}
\label{eq:ff1}
\max \; f_1(x) = \text{spars}(x, A)- k\;\text{dist}(x.m_1, x.m_2)
\end{equation}

\noindent
where $A$ is the archive and $x.m_1, x.m_2$ are the members of the pair in the individual $x$.
Both functions \textit{spars} and \textit{dist} report a measure of distance between inputs. Hence, the constant $k$ is a pure number that can be safely set to $1$. It can also be experimentally tuned (by decreasing it to values less than $1$), to give more importance to the sparseness component of $f_1$, especially when preliminary runs of the algorithm show that the final archive contains a small number of individuals.

Function \textit{dist} measures the similarity of the inputs within an individual as the distance between its members. This distance is computed on the input instances and is domain-specific. For the image classification problem, we compute the Euclidean distance between  pixel matrices~\cite{PeiCYJ17, GuoJZCS18}. For the regression problem, we use a weighted Levenshtein distance \cite{LevenshteinSov66} that takes into account the edit operations on the sequences of angles and points sampled on the spines of the  roads being compared. This distance metric is suitable for the comparison of road shapes, since it takes into account the order of the points along the curves as well as the relative angle between consecutive points.  Function \textit{spars} measures the sparseness of an individual, i.e., its minimum distance from the solutions in the archive $A$.
\text{spars}$(x)$ is defined as the distance from the closest individual in the archive $A$: $\min_{y \in A}\text{dist}(x, y)$. \textit{dist}$(x, y)$ is computed from the distances between  individual members, as the minimum between (\textit{dist}($x.m_1, y.m_1$) + \textit{dist}($x.m_2, y.m_2$))/2 and (\textit{dist}($x.m_1, y.m_2$) + \textit{dist}($x.m_2, y.m_1$))/2.We introduced the sparseness in the quality measure to promote diversity within the solution. This was motivated by preliminary experiments we ran in which the search tended to get stuck in local optima, covering only a tiny portion of the frontier, if sparseness was not included in the fitness function.

\subsubsection{Closeness to the frontier}
The fitness function $f_2$ measures the closeness of an individual to the frontier: 

\begin{equation}
\label{eq:closeness}
\min \; f_2(x) = \begin{cases}
 \text{eval}(x.m_1) \cdot \text{eval}(x.m_2),& \text{if } >0\\
-1,& \text{otherwise}\\
\end{cases}
\end{equation}

Computation of $f_2$ requires the execution of the DL system under test on the two members belonging to the individual. This is represented by the invocation of function \textit{eval} on each member. The quality of the behaviour exhibited by the DL system under test is measured during its execution. We design $f_2$ so that such quality is a positive number if the system exhibits the expected behaviour;  a negative number otherwise.

The definition of function \textit{eval} is domain/problem specific. For the image classification problem, we exploit the confidence level provided by the output layer of the DNN as \textit{eval} function. In fact, the output of a classification DNN is usually the array returned by the softmax activation function containing the confidence levels assigned to each of the possible classes~\cite{GoodfellowMIT16}. More specifically, \textit{eval} is calculated as the difference between the confidence level associated with the expected label and the maximum confidence level associated to any other class.

For the steering angle prediction problem, we use a metric similar to that proposed by Gambi et al. \cite{GambiMF19}. The behaviour of the DNN is characterised by the distance of the car from the center of the lane during the simulation of the corresponding input scenario. More specifically, \textit{eval} is  calculated as $\min (w/2-d)$, where $w$ is the width of the lane and $d$ the distance of the car from the lane centre. The position of the car is approximated by its centre of mass. Function \textit{eval} returns its maximum value ($w/2$) when the car has distance zero from the center of the lane; it returns a negative number when an out of bound episode occurs.

\subsection{Initial Population}

Function \textsc{InitialisePopulation} (line 3 in \autoref{algo}) returns the initial population given a set of seeds and the population size.
Seeds are inputs on which the DL system under test exhibits a correct behaviour. The two members of an individual are obtained by copying the same seed twice and applying the mutation operator to one of the two copies.

More specifically, for image classification, seeds are chosen from the MNIST samples that are correctly classified by the system under test. For steering angle prediction, we generate valid roads and evaluate the system on them. The ones on which the car does not depart from the lane are considered as seeds (positive \textit{eval}). 

\subsection{Archive of Solutions}

The best (non dominated) individuals encountered during the search are kept in the archive~\cite{DeJong04}. This prevents the search for novelty from \textit{cycling}, a phenomenon where the population moves from one area of the behavioural space to another and back again, without any memory of the areas it has already explored~\cite{Mouret15}. At the end of the exploration, the archive  contains the final solution.

The archive is managed by the \textsc{UpdateArchive} function (lines 5 and 15 of \autoref{algo}).
An individual of the population is considered as a candidate to be included in the archive if it is at the frontier of  behaviours. When a new input pair at the frontier is found, it is compared with its nearest neighbour in the archive.
If the distance from the nearest neighbour in the archive is higher than a threshold $t_a$, the new individual is kept in the archive. Instead, if the distance from its nearest neighbour is lower than the threshold, the new candidate competes locally with its nearest neighbour in the archive. The local competition rewards individuals outperforming the most similar ones in the behavioural space~\cite{LehmanS11b}. In our case,  local competition is based on the distance between the members of each pair: only the individual that has the closest members is kept in the archive. 

The threshold $t_a$ is a parameter that determines the granularity of the final frontier, so it can be adjusted by the tester so as to obtain a frontier with the desired size: a high value of $t_a$ makes it difficult for new individuals at the frontier to enter the archive, because they must be extremely different from those already included. Lowering the value of $t_a$ increases the granularity of the frontier and a higher number of similar individuals are added. To choose empirically the value of $t_a$, we recommend to (1) compute the minimum distance among a randomly selected set of diverse inputs; (2) choose a value greater than this number; (3) iteratively adjust this value based on the final archive size.

\subsection{Selection Operator}

We use the \textsc{Select} operator from \textsc{NSGA-II} (lines 6, 9 and 16 in \autoref{algo})~\cite{DebAM02}. This selection operator favours individuals with smaller non-domination rank and, when the rank is the same, i.e., the individuals belong to the same Pareto front, it favours the one with higher crowding distance (less dense regions) to promote diversity. Since we use tournament selection, the offspring of the current population is obtained by choosing the winner among the (two) individuals being compared in each tournament (\textsc{SelTourDCD} at line 9 in \autoref{algo}).

\subsection{Mutation}

The  individuals selected for the offspring are mutated by the \textsc{Mutate} operator (lines 10:12 in \autoref{algo}). This operator manipulates the control parameters of the model representation of each input: it chooses one of the two members of the individual and it applies a perturbation to the model parameters representing the input. The extent of the perturbation is uniformly sampled in a customisable range. 

After applying the operator, we verify if the mutant complies with the constraints of the input domain. Moreover, we also verify that, once concretised into an actual input for the DL system, the mutant is different from its parent and from the other member of the pair. If any of these checks fails, the operator is applied repeatedly, until a valid input is obtained.

For the classification problem, the mutation operator randomly chooses a start point, an end point or a control point of the SVG model and applies a displacement to it in the two-dimensional space. The mutation operator for the regression problem is  similar and it is applied to the control points that define the road shape.

\subsection{Repopulation Operator}

The exploration could get stuck in local optima, despite the mechanisms used to promote diversity (e.g., sparseness, in fitness function$f_1$). To mitigate this undesirable situation and further vary the population, \tool uses the \textsc{Repopulation} operator (line 13 in \autoref{algo}) inspired by the Shotgun hill climbing meta-heuristic algorithm~\cite{HarmanMAZ12}. It replaces \textit{n} of the most dominated individuals in the current population with individuals newly generated from the seeds. The aggressiveness of this operator can be tuned by setting the range from which the value of \textit{n} is uniformly sampled.

When repopulation is applied, each new individual is generated starting from one of the seeds that have not yet produced any solutions in the archive. If all the starting seeds have produced at least one solution, the new individuals are generated starting from a randomly chosen seed. 

\section{Experimental Evaluation}
\label{experiment}
\subsection{Subject Systems}

We evaluate \tool on two DL systems, addressing different tasks and domains.
The first system performs a classification task, which consists of recognising handwritten digits from the MNIST dataset~\cite{LecunBBH98}. 
The second system solves a regression problem. Specifically, it predicts the steering angle of a self-driving car given the image of its onboard camera~\cite{BojarskiNVIDIA16}, using the BeamNG simulator~\cite{beamng_research}. Hereafter, we refer to such objects of study simply as \textit{MNIST} and \textit{BeamNG}. 
We chose these problems because of their representativeness, and because they have been widely used in the literature to evaluate testing techniques for DNN systems~\cite{GuoJZCS18,KimFY19,PeiCYJ17,GambiMF19}. Moreover, self-driving cars are an example of safety-critical usage of DNNs.
To assess the usefulness of the frontier when computed for a high quality (HQ) vs a low quality (LQ) DL system, we trained two versions of each of the two considered systems.

The \textbf{MNIST} case study consists of a DNN model that predicts which digit is represented by an input image. We considered the deep convolutional network provided by Keras,\footnote{\url{https://github.com/keras-team/keras/blob/master/examples/mnist_cnn.py}} because of its popularity, simplicity and effectiveness. \tool crosses the frontier of MNIST when the input image is misclassified. The HQ version of MNIST has $99.11\%$ test accuracy and was obtained by training the DNN on the $60\,000$ images of the MNIST training set with the default settings provided by Keras, i.e., $12$ epochs, with batches of size $128$, and with a learning rate equal to $1$. The LQ version was trained on the same training set and with the same hyper-parameters, with the exception of the learning rate that was set to $0.001$. In fact, such a learning rate is associated with an accuracy drop, i.e., the model's test accuracy goes down to $84.34\%$ .

The \textbf{BeamNG} case study is a self-driving car equipped with a Lane Keeping Assist System (LKAS) running in the BeamNG simulator~\cite{beamng_research}.
It adopts a \textit{behavioural reflex} approach, i.e., the DL component learns a direct mapping from the sensor camera input to the steering angle value to be passed to the actuators~\cite{ChenSKC15}. 
The DNN driving the BeamNG ego-car utilises the \textsc{Dave-2} architecture designed by Bojarski et al. at \textsc{NVIDIA}, consisting of three CNNs, followed by five fully-connected layers~\cite{BojarskiNVIDIA16}.
The DNN was trained with images captured by the camera sensors of the ego-car, paired with the steering angles provided by the simulator's autopilot, which takes advantage of global knowledge and computes the optimal steering angle geometrically. \tool crosses the frontier of BeamNG if the ego-car goes out of bound when driving in the input road. For BeamNG we also produced two versions, HQ and LQ. Both versions were trained for $4,600$ epochs, with batches of size $128$ and with a learning rate equal to $0.001$. To train the LQ version, we used a training dataset obtained by letting the autopilot drive up to 15 mph on a sinusoidal road ($y = sin(x/10) \times 10$, where 1 unit corresponds to 1 meter). The HQ version was instead trained on an enriched training set, including $30$ diverse types of roads made of 20 control points that were automatically generated. Unrepresentative training data is a common fault in DL systems~\cite{HumbatovaICSE20}.

\subsection{Research Questions}

\textbf{RQ1 (Effectiveness):} \textit{What is the intersection between the frontier reported by \tool and the input validity domain of the DL system under test?}

This is the main research question of our empirical evaluation, since it focuses on the use of \tool to check if the frontier behaviours of the DL system under test are outside the validity domain, which  indicates the system has reached an adequate quality level, or within the validity domain, which points to issues that might affect the DL system in real executions. 

\textbf{Metrics:} Assessing whether a frontier input is or is not part of the valid inputs is in general a domain dependent task, which requires human judgment and deep knowledge of the requirements behind the DL system. For MNIST, we resort to a human study in order to understand if and to what extent a given digit is recognisable correctly by a human. When this is not the case, the input  image is deemed as outside the validity domain of the classifier. For BeamNG, we refer to the guidelines from the American Association of State Highway and Transportation Officials (AASHTO)~\cite{aashto2018}, which prescribes among other things the minimum recommended radius of curvature by speed limit (in particular, 47 ft at 15 mph, the car speed in the simulations run on BeamNG). When the generated inputs violate the road design guidelines on the minimum curvature radius we deem the frontier input as outside the validity domain. 

\textbf{Human assessment:} To determine whether the frontier of MNIST intersects its validity domain, we asked humans to recognise the digit images taken from the frontier and to declare their confidence in such recognition.  
We created $20$ surveys made of $10$ questions to be presented to human assessors: $9$ assessment questions (ASQ) and $1$ attention check question (ACQ). The difference between ASQ and ACQ is that the former involves a frontier image while the latter is by construction a nominal image whose classification is trivial and unambiguous for humans. To this aim, we used the digits delivered with the Freestyle Script font,\footnote{\url{https://www.fonts.com/font/itc/freestyle-script}} a computer font that resembles handwritten characters. We double checked manually that the ASQs were indeed unambiguous to classify.
To obtain the images featured in the ASQs, we ran \tool on both versions of MNIST (HQ and LQ), once for each digit class. Then, we considered the member of the pair outside the frontier (i.e., the   misclassified image). We took $90$ images for each of the two MNIST frontiers. We sorted the elements in the frontier based on their distance from the corresponding nominal Freestyle-font digit (the same image used in the ACQ) and divided them into $9$ buckets of the same size. We selected the same number of samples ($9$) for each digit by randomly selecting one sample from each bucket. In total, we selected $180$ inputs from the obtained frontiers. Each of the selected inputs appeared in a single survey. The order of appearance of HQ ASQs, LQ ASQs and the ACQ was randomised in the surveys.  In total, this assessment involved $20$ different human assessors, one for each survey.

For each image, we asked humans to answer the following questions: (1) \textit{``What digit does the image represent?'' (0, 1, 2, $\ldots$, 9)}; (2) \textit{``How confident are you in your answer?'' (-2: not at all, -1: not much, 0: borderline, 1: quite confident, 2: very confident)}.

\textbf{RQ2 (Discrimination):} \textit{Does \tool provide discriminative information about the DL systems under test?}

In this research question, we compare the frontiers of two DL systems, one exhibiting good performance (HQ) and one having poor performance (LQ). We measured the size of the region identified by each frontier, to assess whether  HQ systems have a larger frontier than LQ ones.

\textbf{Metrics:} To answer this research question quantitatively, we defined the metric \textit{radius}: let us consider the archive $A$ at the end of the execution of \autoref{algo} and let us assume that each individual $x \in A$ stores  the input on which the DL system misbehaves in its second member $x.m_2$. We define the outer frontier of misbehaviours as $S_{out} = \{ x.m_2 | x \in A\}$ and the inner one as $S_{in} = \{ x.m_1 | x \in A\}$. The radius measures the average distance of inputs in the frontier from the reference input $\Omega$, an elementary, nominal input that the system is expected  to handle correctly by the requirements: 

\begin{equation}
\label{eq:radius}
\text{radius}(S) = \frac{\sum_{m \in S} \text{dist}(m, \text{$\Omega$})}  {|S|}
\end{equation}

For MNIST, we considered as reference image $\Omega$ the corresponding digit in Freestyle Script font, converted to the same format as the MNIST dataset. For BeamNG, the reference sample is a straight road with no curves. 

We also evaluated  HQ's vs LQ's frontiers qualitatively, by involving humans in a survey, in which they performed pairwise comparisons between images taken from the two frontiers (i.e., one image from HQ's and one from LQ's frontier).

\textbf{Human assessment:} 
For the qualitative assessment of HQ's vs LQ's frontiers, we provided human evaluators with two images taken respectively from each of the two frontiers. For MNIST, we asked them to decide which of the two digit images is easier to recognise. For BeamNG, we asked them to decide which of the two roads is easier to drive. Each pair of inputs was assessed by two human evaluators. Then, we measured the number of answers in which both subjects agreed in considering as  easier to recognise/drive  the element on the frontier of HQ (resp. LQ). 
This would support our conjecture that manual inspection of the frontier is an effective way to discriminate between good and poor performance DL systems. 
For each system, we published  $10$ surveys made of $10$ questions: $9$ discriminative questions (DSQ) and $1$ attention check question (ACQ), consisting of a pair of inputs for which the human choice is obvious and completely predictable. 
Each survey, made of $10$ questions, was answered by $2$ evaluators. In total, this assessment involved $40$ different evaluators, i.e., $20$ for each system.

\textbf{RQ3 (Comparison):} \textit{Is \tool able to characterise the frontier of the behaviours better than the state of the art tool \textsc{DLFuzz}?}

\textsc{DLFuzz}~\cite{GuoJZCS18} is a state of the art tool for the generation of boundary values by means of fuzzing. Among the techniques proposed in the literature, DLFuzz is the most related to \tool since it produces boundary inputs by manipulation of existing seeds. DLFuzz uses gradient ascent optimisation  to maximise a custom loss function that takes also into account  the distance between the new input and the seed. It should be noticed that DLFuzz is not a model-based input generator as it operates directly on the raw input (i.e., image pixels). Hence, its boundary inputs are supposedly less realistic and less representative than \tool'. 

\textbf{Metrics:} We compare the radius, as defined above, of \tool' frontier w.r.t. DLFuzz's boundary inputs.

\subsection{Experimental Procedure}

Our experimental procedure consists of: (1)  generation of the frontiers for the DL systems under test (MNIST HQ/LQ; BeamNG HQ/LQ) and computation of the radius for the generated frontiers; (2) generation of boundary inputs using DLFuzz and comparison with \tool's frontiers; (3) human assessment of the inputs in the frontiers, by means of two surveys: a digit recognition survey (for RQ1) and a pairwise image comparison survey (for RQ2).

\textbf{Generation of the frontier with \tool:}
We ran \tool $10$ times on each version of each system under test. At the end of the runs, we collected the values of the radius metric, as well as the representation of the input pairs that belong to the frontiers. The configurations of \tool were obtained in a few preliminary runs and are reported in \autoref{tab:config}. 

Before each run, we obtained a different set of initial seeds by sampling the inputs under the constraint that they have to produce a correct behaviour of the systems under test. For MNIST, each set of seeds was obtained by randomly selecting $100$ correctly classified inputs from the MNIST test set, all belonging to the same class (digit ``5''). Similar results have been obtained for digits other than five, but we do not report them for space reasons. For BeamNG, each initial set consisted of $12$ valid seed roads on which the considered model was able to keep the lane. A seed road was defined by $10$ control points in which the initial point was always at a fixed position whereas the others were placed at a random position 25 meters away from the previous one.

\begin{table}
\setlength{\tabcolsep}{8pt}
\renewcommand{\arraystretch}{1.1}
\centering
\caption{DeepJanus Configurations}

\begin{small}
\begin{tabular}{ l r r}

\toprule

 Parameter & MNIST & BeamNG \\ 
 
 \midrule
 
 population size & 100 & 12 \\  
 generations & 4000 & 100 \\
 mutation lower bound & 0.01 & 1 \\
 mutation upper bound & 0.6 & 6 \\
 archive threshold $t_a$ & 4 & 35 \\
 repopulation upper bound & 10 & 2 \\
 parameter $k$ of fitness function $f_1$ & 0.1 & 0.01 \\
 
 \bottomrule
 
\end{tabular}
\end{small}
\label{tab:config}
\end{table}

\textbf{Generation of boundary inputs with DLFuzz:} We generated boundary inputs for MNIST using \textsc{DLFuzz}~\cite{GuoJZCS18} starting from the same seeds used by \tool. However, we could not consider BeamNG in the comparison, because  \textsc{DLFuzz} is not able to test a system in the simulation loop, on a sequence of images.  It can only evaluate the output of the  DNN component on a single, statically collected image, which is fuzzed by the tool. This means that it is not possible to simulate out of bound episodes, which would require a sequence of images to be fuzzed dynamically. We adopted the \textsc{DLFuzz} configuration that is reported as the one achieving the best performance~\cite{GuoJZCS18}.

\textbf{Human assessment of the frontier:} We outsourced our surveys to a crowdsourcing platform in order to have a diverse pool of respondents~\cite{BehrendSMW11}. \textit{Crowdsourcing} has recently become quite popular in software engineering~\cite{MaoCHJ17} to automate tasks that can only be performed by humans. A problem is specified  in the form of small Human Intelligence Tasks (HITs) and made available in a crowdsourcing platform, where registered workers can choose to complete HITs for a small remuneration~\cite{PastoreMF13}. We selected the Amazon Mechanical Turk platform\footnote{\url{https://www.mturk.com}} for our two surveys (resp. for RQ1 and RQ2), because it is well known, well documented and widely used to gather qualitative feedbacks~\cite{KitturCS08, HeerB10}. 

We applied two methods to ensure the quality of the answers: (1) added  an attention check question (ACQ) to each survey (see above); and (2) restricted the participation to workers with high reputation (above 95\% approval rate)~\cite{PeerVA14}. We only accepted answers from users that passed the ACQ. 

\section{Results}
\label{result}
 
\subsection{RQ1 (Effectiveness)}

\begin{table}
\setlength{\tabcolsep}{8pt}
\renewcommand{\arraystretch}{1.1}
\centering
\caption{RQ1: Invalid inputs found at the frontier. The metrics in the last column are different since the validity domains of MNIST and BeamNG are assessed through system specific metrics, i.e. human assessment + confidence and curvature radius, respectively}


\begin{small}
\begin{tabular}{ l l r r}

\toprule

 Object & Validity & Number & Confidence\\ 
 
 \midrule

 \multirow{2}{*}{MNIST HQ} & Valid & 69 & 0.463 $\pm$ 1.255 \\  
 & Invalid & 21 & -0.095 $\pm$ 1.338 \\  
 \multirow{2}{*}{MNIST LQ} & Valid & 82 & 1.524 $\pm$ 0.835 \\  
 & Invalid & 8 & -0.875 $\pm$ 1.356\\

 \midrule
 
 $p$-value & &1.394E-2 & 2.48E-4 \\
 odds ratio & & 0.322 & - \\
 effect size & & - & 0.81 (large) \\ 
 
 \bottomrule

\toprule

 Object & Validity & Number & Curv. radius (ft)\\
 
 \midrule
 
 \multirow{2}{*}{BeamNG HQ} & Valid & 1 & 47.371 \\  
 & Invalid & 141 & 37.359 $\pm$ 4.262 \\  
 \multirow{2}{*}{BeamNG LQ} & Valid & 56 & 49.283 $\pm$ 1.966 \\  
 & Invalid & 173 & 40.419 $\pm$ 4.488\\

 \midrule
 
 $p$-value & &3.787E-12 & 2.52E-12 \\
 odds ratio & & 0.022 & - \\
 effect size & & - & 1.009 (large) \\ 
  
 \bottomrule
 
\end{tabular}
\end{small}

\label{tab:rq1}
\end{table}

\autoref{tab:rq1} reports the intersection between the input validity domain and the outer frontier of behaviours for each version of the considered systems. The upper part of the table reports the results for the MNIST system. As shown in the first two rows, out of the $90$  images generated by \tool on MNIST HQ, $69$ are recognised by the crowdworkers as the digit classes to which the corresponding seeds belong, whereas $21$ are recognised incorrectly. On the other hand, on MNIST LQ \tool generated  $82$ frontier inputs that are recognised correctly by crowdworkers. This indicates that the frontier of MNIST LQ has a larger intersection with the set of valid inputs than MNIST HQ. The classification performed by the crowdworkers was subjected to the Fisher's exact test~\cite{Fisher06} to determine the statistical significance of the effect of the version (HQ vs LQ) on the validity of the frontier inputs. The $p$-value lower than the usual threshold $\alpha = 0.05$ indicates statistical significance of the difference between MNIST HQ and MNIST LQ. The odds ratio indicates that the expected relative proportion of valid vs invalid is much lower in the MNIST HQ system.

The last column contains the confidence (mean $\pm$ standard deviation) expressed by the crowdworkers when classifying the images. The scale is between -2:+2 (with -2 = min confidence and +2 max confidence). Human assessors had a significantly higher confidence in classifying the valid inputs belonging to the frontier of MNIST LQ and were more uncertain when recognising valid inputs belonging to the frontier of MNIST HQ. This confirms that the frontier of a high quality DL system contains elements that are difficult to classify confidently even for a human. We assessed the statistical significance of the confidence comparison by applying Generalised Linear Modelling (GLM)~\cite{NelderStat72}. The dependent variable of the GLM model is confidence, whereas the independent variable is a numeric encoding of the system (HQ = 0; LQ = 1). The results of the statistical test are a small $p$-value (way below $\alpha = 0.05$) and a large Cohen-d effect size (i.e., a large difference between the means, normalised by the pooled standard deviation; conventionally, the threshold for a large Cohen-d effect size is set to 0.8).

The lower part of the table shows the results for the BeamNG system. As reported in the third column, the self-driving car equipped with the HQ lane keeping assist system goes out of bound (misbehaves) only on one valid road from the frontier (indeed the minimum radius of curvature of this road is greater that the AASHTO threshold by just 0.371 feet). Instead, there is a significant number of valid frontier roads when the car is equipped with the LQ system. The fourth column reports the minimum radius of curvature (mean $\pm$ standard deviation). Its values for the roads on the HQ frontier are significantly lower than the radius values on the LQ frontier. This confirms that roads in the HQ frontier are substantially more difficult or even impossible to drive than those in the LQ frontier. Also for the results of BeamNG, we applied the Fisher's exact test and GLM to assess the statistical significance of the results. $P$-values indicate statistical significance of the difference between HQ and LQ, with a low odds ratio (valid/invalid proportion in HQ vs LQ) and a large Cohen-d effect size (large normalised difference between minimum curvature radii).

\begin{tcolorbox}
\textbf{Summary}: \textit{Many elements at the frontier of behaviours identified by \tool intersect the input validity domain when the quality of the system under test is low. For a high quality system, the valid inputs at the frontier are challenging to handle even for humans (MNIST case study) or are very close to being invalid (BeamNG case study).} 
\end{tcolorbox}

\subsection{RQ2 (Discrimination)}

\begin{table}[t]
\setlength{\tabcolsep}{5pt}
\renewcommand{\arraystretch}{1.1}
\centering
\caption{RQ2 (top, middle): discriminating HQ from LQ by \tool (DJ)'s inner/outer radius and by DLFuzz (DLF); RQ3 (bottom): comparing  \tool vs DLFuzz }
\label{tab:rq2} 


\begin{small}
\begin{tabular}{lllrrrr}

\toprule

& &  & Inner Radius & Outer Radius\\
 
 \midrule

\multirow{8}{*}{DJ} & \multirow{4}{*}{MNIST} & HQ & 10.575 $\pm$ 0.188 & 10.597 $\pm$ 0.188 \\

& & LQ & 10.284 $\pm$ 0.083 & 10.328 $\pm$ 0.078\\
 
& & $p$-value & 2.96E-4 & 5.61E-4\\
  
& & effect size & 1.99 (large) & 1.87 (large) \\ [0.5ex] 
 
& \multirow{4}{*}{BeamNG} & HQ & 55.968 $\pm$ 1.522 & 57.236 $\pm$ 1.753\\
 
& & LQ & 52.853 $\pm$ 2.22 & 55.565 $\pm$ 2.412 \\
 
& & $p$-value & 1.8E-2 & 9.34E-2\\
  
& & effect size & 1.636 (large) & 0.792 (medium)\\ [0.5ex]

 \midrule

\multirow{4}{*}{DLF} & \multirow{4}{*}{MNIST} & HQ & - & 9.889 $\pm$ 0.275 \\

& & LQ & - & 9.976 $\pm$ 0.115\\

& & $p$-value & - & 0.385\\
  
& & effect size & - & -0.398 (small) \\ [0.5ex] 

 \midrule

\multicolumn{2}{l}{\multirow{2}{*}{DJ vs DLF (HQ) }} & $p$-value & - & 4.14E-6 \\
\multicolumn{2}{l}{} & effect size & - & 2.905 (large) \\

\multicolumn{2}{l}{\multirow{2}{*}{DJ vs DLF (LQ) }} & $p$-value & - & 4.25E-7 \\
\multicolumn{2}{l}{} & effect size & - & 3.441 (large)\\

\bottomrule
 
\end{tabular}
\end{small}

\end{table}

The top rows of \autoref{tab:rq2} show the inner/outer frontier radii returned by \tool for the systems under test (mean $\pm$ standard deviation).

For each system under test, we compare the values of the frontier radii obtained for HQ vs LQ, to understand if the frontier volume can help discriminate between different quality levels. We assessed the statistical significance of this comparison by applying GLM, with radius as dependent variable and the quality of the system (HQ = 0; LQ = 1) as independent variable. 

For MNIST, the radius for the LQ version of the system is significantly smaller than the one of the HQ version (with low $p$-value and large Cohen-d effect size). This means that, on average, the higher quality system tolerates larger changes to input images before exhibiting a misbehaviour. Similarly, for BeamNG the radius for the LQ version of the system is significantly smaller than the one of the HQ version (with low $p$-value and large/medium Cohen-d effect size resp. for inner/outer radius), showing again that the frontier can discriminate the HQ version from the LQ one.

\begin{table}[t]
\setlength{\tabcolsep}{7pt}
\renewcommand{\arraystretch}{1.1}
\centering
\caption{RQ2: human discrimination of LQ from HQ based on frontier inputs (MNIST: easier to recognise digit; BeamNG: easier to drive road)}


\begin{small}
\begin{tabular}{ l r r r r r}

\toprule

 Object & LQ & HQ & disagree & $p$-value & $p$-success \\ 
 
 \midrule
 
 $MNIST$ & 77 & 4 & 19 & < 2.2E-16 & 0.95 \\  
 $BeamNG$ & 79 & 9 & 12 & 4.152E-15 & 0.89 \\
 
\bottomrule
 
\end{tabular}
\end{small}
\label{tab:rq2a}
\end{table}

\autoref{tab:rq2a} shows the results  obtained from the crowdsourced survey on the discrimination of easier to recognise digits and easier to drive roads. In the LQ column, it reports the number of answers in which both crowdworkers comparing the same pair of images agreed in considering as qualitatively easier to classify/drive the input on the frontier produced by the LQ system. The HQ column reports the number of answers in which both  subjects agreed in considering as qualitatively easier to classify/drive the input on the HQ frontier. The next column reports the number of cases in which the two subjects did not agree with each other. 

Crowdworkers were able to determine the relative quality of the systems by looking at the inputs on their frontiers with very high accuracy. In fact, the large majority (95\% for MNIST and 89\% for BeamNG) of inputs found by \tool on the outer frontier of the LQ system are perceived as qualitatively easier to classify/drive than those of the HQ systems. Statistical significance of the classification performed by the users was assessed by applying the Binomial exact test~\cite{ClopperBiom34}. The returned $p$-values for MNIST and BeamNG are very low, showing that the choice between LQ frontier input and HQ frontier input as the easier to classify/drive is very unlikely to be random and uniform (i.e., 50\%, 50\%).

\begin{tcolorbox}
\textbf{Summary}: \textit{The frontier of behaviours  allows  developers to discriminate a system with higher from one with lower quality: the  radius is significantly larger for the former. A larger frontier means that the system is able to generalise correctly the learned behaviour to a larger set of valid input data. This is confirmed by the human study, where inputs from the larger frontier were deemed to be more difficult to handle than those from the smaller frontier.}
\end{tcolorbox}
 
\subsection{RQ3 (Comparison)}

In order to address $RQ_3$, we compare the frontier identified by our approach with the boundary inputs returned by DLFuzz~\cite{GuoJZCS18}. The middle part of \autoref{tab:rq2} shows the radius of the outer boundary inputs returned by DLFuzz for the MNIST classifier (inner boundaries are not returned by DLFuzz). We can notice that, differently from the frontier produced by \tool, the boundary inputs of DLFuzz do not exhibit any statistically significant difference between HQ and LQ's frontier radius (high $p$-value, $>$ 0.05). Actually, the difference between the radii seems to go in the opposite direction than in the case of \tool (negative, small effect size). This indicates that DLFuzz cannot reliably discriminate HQ from LQ by generating a larger frontier for the former than for the latter.

As reported in the lower part of \autoref{tab:rq2}, for both LQ and HQ the radius of the frontiers identified by DLFuzz is always significantly smaller than the radius of \tool. 
This indicates that \tool can explore the frontier more thoroughly than DLFuzz.

Indeed, DLFuzz applies small perturbations to the pixels of the input image that cause the classifier to fail, but it does not attempt to explore the input space thoroughly by promoting diversity when new inputs are generated (the only way to promote diversity in DLFuzz is by selecting different initial seeds at each run, which we did in our experiments). On the contrary, \tool is guided by a fitness function ($f_1$) that accounts explicitly for the sparseness of the generated inputs, hence exploring the behavioural space more thoroughly. As a consequence, the inputs generated by DLFuzz are closer to the reference than the ones produced by \tool.

\begin{figure}
\centering
  \includegraphics[width=0.9\linewidth]{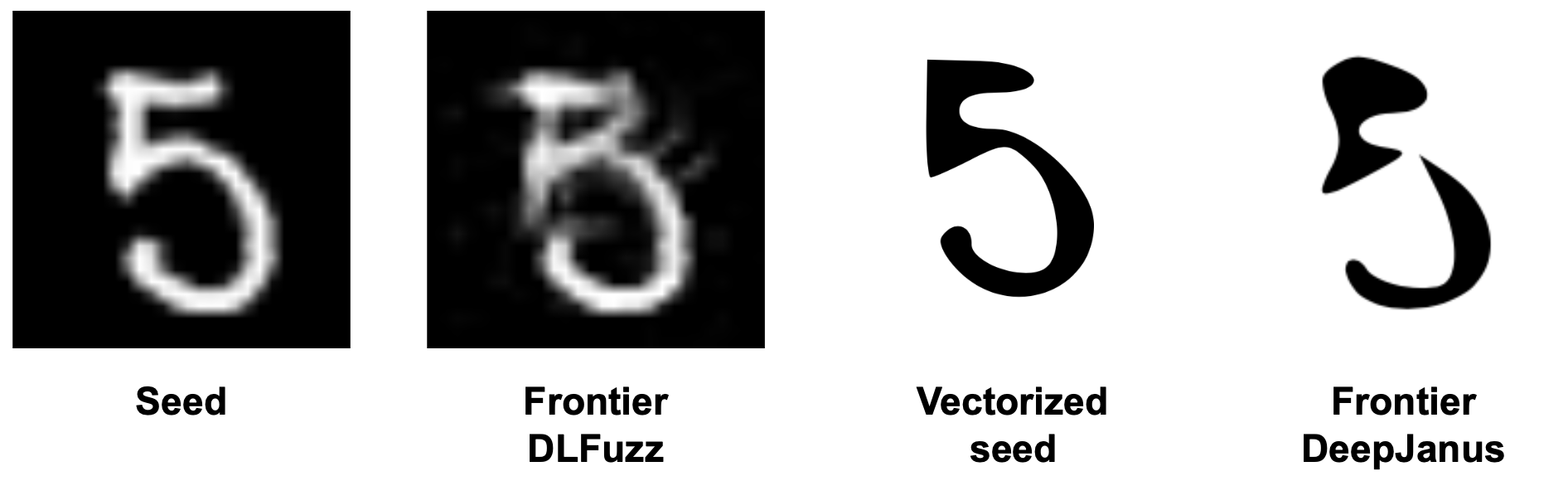}
  \caption{Seed and frontier images generated respectively by \tool and DLFuzz (the starting seed is the same)}
  \label{fig:dlfuzz}
\end{figure}

\begin{figure}
\centering
  \includegraphics[width=\linewidth]{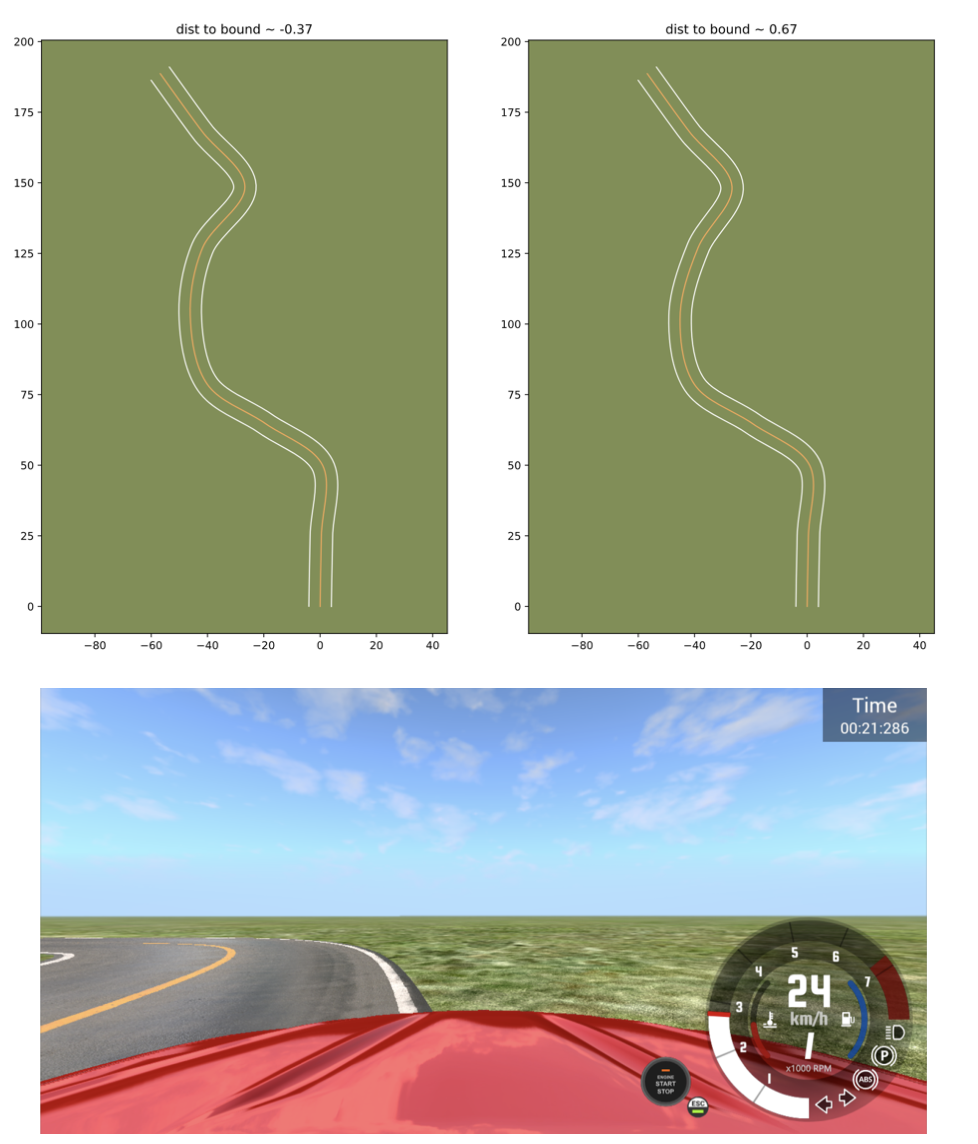}
  \caption{A pair of inputs at the frontier of the BeamNG system (top). The left input triggers a misbehaviour, i.e. the car goes out of bounds, as depicted in the bottom picture
  }
  \label{fig:frontier_BNG}
\end{figure}

From a qualitative point of view, the boundary inputs found by DLFuzz are quite different from those found by \tool, as apparent from Fig.~\ref{fig:dlfuzz}. In fact, the same seed is manipulated as a shape of calligraphic traits modelled in SVG by \tool (two rightmost pictures in Fig.~\ref{fig:dlfuzz}), while raw pixel manipulation is performed by DLFuzz (leftmost pictures). The images generated by \tool simulate quite  realistically the traits that can be found in a hand-written digit, while DLFuzz's images look like blurred, noisy versions of the original ones.

As regards the steering angle prediction task, we cannot perform a direct comparison with existing techniques. In fact, existing model-based techniques ~\cite{AbdessalemNBS16, AbdessalemPNBS18, AbdessalemNBS18, GambiMF19} can find system failures but do not aim at finding frontier inputs (as shown at the top of \autoref{fig:frontier_BNG}). Instead, techniques that perform raw data manipulation ~\cite{PeiCYJ17, GuoJZCS18, TianPSB18, ZhangZZ0K18} just manipulate individual camera images and test the steering angle predictor in isolation. Thus, they cannot assess whether an incorrect prediction causes an out of bound episode, as the one shown at the bottom of \autoref{fig:frontier_BNG}.

\begin{tcolorbox}
\textbf{Summary}: \textit{\tool explores a significantly larger frontier than DLFuzz. The images produced by \tool for MNIST look more realistic than those of DLFuzz.}
\end{tcolorbox}

\subsection{Threats to Validity}

\textbf{Construct Validity}: The choice of the reference input, $\Omega$, for the computation of the radius in a given domain is conventional. We have tried different choices of $\Omega$ in both domains (MNIST and BeamNG) and found that the experimental results reported in \autoref{result} are not sensitive to the specific choice of $\Omega$: the same conclusions were drawn when a different choice of $\Omega$ was made. The choice of the distance metrics is also crucial. We did not fine tune the metrics for the case study domains, but we adopted metrics that have been used in previous studies from the literature, i.e. we used Euclidean distance when comparing matrices of grayscale values (MNIST) and Levenshtein distance when comparing sequences of road points (BeamNG). Finally, the eval function called within our fitness function $f_2$ is also domain specific. Also in this case, we chose sound metrics that have been already adopted in the literature, i.e. confidence of digit classification and distance from the center of the road.

\textbf{External Validity}: The choice of subject DL systems is a possible threat to the \textit{external validity}. To mitigate this threat, we chose two diverse DL systems. One is a DNN that solves a classification problem, while the other  is a self-driving car equipped with a DNN component that solves a regression problem. However, our results might not generalise to other DL systems and further studies with a wider set of systems should be carried out to fully assess the generalisability of our findings. 

To ensure \textbf{Reproducibility} of our results, the source code of \tool, our objects, and the experimental data are available online~\cite{tool}, making the evaluation repeatable.

\section{Related Work}
\label{related}
\textbf{Raw Input Data Manipulation:}
Several works generate test inputs for DL systems by applying perturbations to the available training/test data. These approaches aim at generating inputs that trigger inconsistencies between multiple DL systems~\cite{PeiCYJ17}, or between the original and a transformed test input~\cite{GuoJZCS18, TianPSB18, ZhangZZ0K18}. They require white-box access to the activation levels of the DNN, if they are guided by coverage criteria such as neuron coverage~\cite{PeiCYJ17} or the more fine grained $k$-multisection neuron coverage~\cite{Ma-ASE-2018}.

One of the limitations of these approaches is the lack of realism of the generated inputs. While these corrupted images are  useful for security testing as adversarial attacks,  they are not representative of data captured by sensors of a real DL system. To generate realistic inputs we adopted a model-based approach, which ensures that data are generated only within the constraints of the model. Another limitation of the existing input generators is that they do not attempt to delimit the region of misbehaviour but they just provide a way to sample it, regardless of the distance from the region of nominal behaviour. On the contrary, we sample the entire frontier of misbehaviours as thoroughly as possible.

Kim et al.~\cite{KimFY19} designed a test adequacy criterion, named \textit{surprise adequacy}, to capture the novelty of the input with respect to the training data. However, it has been used for test case selection and retraining but not for test input generation. Moreover, a surprise adequate test set is not necessarily one that covers also our frontier of misbehaviours, so we view the two approaches as complementary.

\textbf{Model-based Input Generation:}
Abdessalem et al.~\cite{AbdessalemNBS16, AbdessalemPNBS18, AbdessalemNBS18} combine genetic algorithms and machine learning to test advanced driver-assistance systems in an industrial setting. Gambi et al.~\cite{GambiMF19} propose \textsc{AsFault}, a search-based approach to test the lane-keeping  system of self-driving cars. The goal of these techniques is to generate extreme and challenging scenarios, maximising the number of detected system failures. \tool is also model-based, but it differs from existing approaches because it aims at reaching the frontier without surpassing it and because it  spreads the generated inputs along the frontier. Its output is not just a set of critical inputs: it is also a thorough characterisation of the frontier of behaviours. 

\section{Conclusions and Future Work}
\label{conclusions}

\tool characterises the quality of a DL system as its frontier of behaviours, i.e., pairs of similar inputs that trigger different behaviours of the system and are far from each other. Experimental results show that frontier inputs provide the developers with both quantitative and qualitative information, useful to assess the quality of a DL system and to identify valid inputs that it cannot handle properly. Our empirical  study  shows that \tool is more effective in characterising the frontier of behaviours than the state of the art tool, DLFuzz. In our future work, we plan to extend the validity of our results by considering a wider sample of DL systems with increasingly complex input domain, including industrial ones.

\section*{Acknowledgements}
This work was partially supported by the H2020 project PRECRIME, funded under the ERC Advanced Grant 2017 Program (ERC Grant Agreement n. 787703). 
The driving simulator has been provided by BeamNG GmbH, and fundamental support has been offered by their employee M. Mueller.
The authors would like to thank N. Brochado, L. Frunzio, and S. Giacomelli for their contributions to the project.

\bibliographystyle{ACM-Reference-Format}
\bibliography{biblio}


\begin{thebibliography}{51}


\ifx \showCODEN    \undefined \def \showCODEN     #1{\unskip}     \fi
\ifx \showDOI      \undefined \def \showDOI       #1{#1}\fi
\ifx \showISBNx    \undefined \def \showISBNx     #1{\unskip}     \fi
\ifx \showISBNxiii \undefined \def \showISBNxiii  #1{\unskip}     \fi
\ifx \showISSN     \undefined \def \showISSN      #1{\unskip}     \fi
\ifx \showLCCN     \undefined \def \showLCCN      #1{\unskip}     \fi
\ifx \shownote     \undefined \def \shownote      #1{#1}          \fi
\ifx \showarticletitle \undefined \def \showarticletitle #1{#1}   \fi
\ifx \showURL      \undefined \def \showURL       {\relax}        \fi
\providecommand\bibfield[2]{#2}
\providecommand\bibinfo[2]{#2}
\providecommand\natexlab[1]{#1}
\providecommand\showeprint[2][]{arXiv:#2}

\bibitem[\protect\citeauthoryear{Abdessalem, Nejati, Briand, and
  Stifter}{Abdessalem et~al\mbox{.}}{2016}]%
        {AbdessalemNBS16}
\bibfield{author}{\bibinfo{person}{Raja~Ben Abdessalem}, \bibinfo{person}{Shiva
  Nejati}, \bibinfo{person}{Lionel~C. Briand}, {and} \bibinfo{person}{Thomas
  Stifter}.} \bibinfo{year}{2016}\natexlab{}.
\newblock \showarticletitle{Testing advanced driver assistance systems using
  multi-objective search and neural networks}. In
  \bibinfo{booktitle}{\emph{Proceedings of the 31st {IEEE/ACM} International
  Conference on Automated Software Engineering, {ASE}}}.
  \bibinfo{pages}{63--74}.
\newblock


\bibitem[\protect\citeauthoryear{Abdessalem, Nejati, Briand, and
  Stifter}{Abdessalem et~al\mbox{.}}{2018a}]%
        {AbdessalemNBS18}
\bibfield{author}{\bibinfo{person}{Raja~Ben Abdessalem}, \bibinfo{person}{Shiva
  Nejati}, \bibinfo{person}{Lionel~C. Briand}, {and} \bibinfo{person}{Thomas
  Stifter}.} \bibinfo{year}{2018}\natexlab{a}.
\newblock \showarticletitle{Testing Vision-based Control Systems Using
  Learnable Evolutionary Algorithms}. In \bibinfo{booktitle}{\emph{Proceedings
  of the 40th International Conference on Software Engineering}} (Gothenburg,
  Sweden) \emph{(\bibinfo{series}{ICSE '18})}. \bibinfo{publisher}{ACM},
  \bibinfo{address}{New York, NY, USA}, \bibinfo{pages}{1016--1026}.
\newblock
\showISBNx{978-1-4503-5638-1}
\urldef\tempurl%
\url{https://doi.org/10.1145/3180155.3180160}
\showDOI{\tempurl}


\bibitem[\protect\citeauthoryear{Abdessalem, Panichella, Nejati, Briand, and
  Stifter}{Abdessalem et~al\mbox{.}}{2018b}]%
        {AbdessalemPNBS18}
\bibfield{author}{\bibinfo{person}{Raja~Ben Abdessalem},
  \bibinfo{person}{Annibale Panichella}, \bibinfo{person}{Shiva Nejati},
  \bibinfo{person}{Lionel~C. Briand}, {and} \bibinfo{person}{Thomas Stifter}.}
  \bibinfo{year}{2018}\natexlab{b}.
\newblock \showarticletitle{Testing Autonomous Cars for Feature Interaction
  Failures Using Many-objective Search}. In
  \bibinfo{booktitle}{\emph{Proceedings of the 33rd ACM/IEEE International
  Conference on Automated Software Engineering}} (Montpellier, France)
  \emph{(\bibinfo{series}{ASE 2018})}. \bibinfo{publisher}{ACM},
  \bibinfo{address}{New York, NY, USA}, \bibinfo{pages}{143--154}.
\newblock
\showISBNx{978-1-4503-5937-5}
\urldef\tempurl%
\url{https://doi.org/10.1145/3238147.3238192}
\showDOI{\tempurl}


\bibitem[\protect\citeauthoryear{Barry and Goldman}{Barry and Goldman}{1988}]%
        {BarryG88}
\bibfield{author}{\bibinfo{person}{Phillip~J. Barry} {and}
  \bibinfo{person}{Ronald~N. Goldman}.} \bibinfo{year}{1988}\natexlab{}.
\newblock \showarticletitle{A Recursive Evaluation Algorithm for a Class of
  Catmull-Rom Splines}.
\newblock \bibinfo{journal}{\emph{SIGGRAPH Comput. Graph.}}
  \bibinfo{volume}{22}, \bibinfo{number}{4} (\bibinfo{date}{June}
  \bibinfo{year}{1988}), \bibinfo{pages}{199--204}.
\newblock
\showISSN{0097-8930}
\urldef\tempurl%
\url{https://doi.org/10.1145/378456.378511}
\showDOI{\tempurl}


\bibitem[\protect\citeauthoryear{{BeamNG GmbH}}{{BeamNG GmbH}}{[n.d.]}]%
        {beamng_research}
\bibfield{author}{\bibinfo{person}{{BeamNG GmbH}}.}
  \bibinfo{year}{[n.d.]}\natexlab{}.
\newblock \bibinfo{booktitle}{\emph{{B}eam{NG}.research}}.
\newblock
\urldef\tempurl%
\url{https://www.beamng.gmbh/research}
\showURL{%
\tempurl}


\bibitem[\protect\citeauthoryear{Behrend, Sharek, Meade, and Wiebe}{Behrend
  et~al\mbox{.}}{2011}]%
        {BehrendSMW11}
\bibfield{author}{\bibinfo{person}{Tara~S. Behrend}, \bibinfo{person}{David~J.
  Sharek}, \bibinfo{person}{Adam~W. Meade}, {and} \bibinfo{person}{Eric~N.
  Wiebe}.} \bibinfo{year}{2011}\natexlab{}.
\newblock \showarticletitle{The viability of crowdsourcing for survey
  research}.
\newblock \bibinfo{journal}{\emph{Behavior Research Methods}}
  \bibinfo{volume}{43}, \bibinfo{number}{3} (\bibinfo{date}{25 Mar}
  \bibinfo{year}{2011}), \bibinfo{pages}{800}.
\newblock
\showISSN{1554-3528}
\urldef\tempurl%
\url{https://doi.org/10.3758/s13428-011-0081-0}
\showDOI{\tempurl}


\bibitem[\protect\citeauthoryear{Bojarski, Testa, Dworakowski, Firner, Flepp,
  Goyal, Jackel, Monfort, Muller, Zhang, Zhang, Zhao, and Zieba}{Bojarski
  et~al\mbox{.}}{2016}]%
        {BojarskiNVIDIA16}
\bibfield{author}{\bibinfo{person}{Mariusz Bojarski},
  \bibinfo{person}{Davide~Del Testa}, \bibinfo{person}{Daniel Dworakowski},
  \bibinfo{person}{Bernhard Firner}, \bibinfo{person}{Beat Flepp},
  \bibinfo{person}{Prasoon Goyal}, \bibinfo{person}{Lawrence~D. Jackel},
  \bibinfo{person}{Mathew Monfort}, \bibinfo{person}{Urs Muller},
  \bibinfo{person}{Jiakai Zhang}, \bibinfo{person}{Xin Zhang},
  \bibinfo{person}{Jake Zhao}, {and} \bibinfo{person}{Karol Zieba}.}
  \bibinfo{year}{2016}\natexlab{}.
\newblock \showarticletitle{End to End Learning for Self-Driving Cars}.
\newblock \bibinfo{journal}{\emph{CoRR}}  \bibinfo{volume}{abs/1604.07316}
  (\bibinfo{year}{2016}).
\newblock
\showeprint[arxiv]{1604.07316}
\urldef\tempurl%
\url{http://arxiv.org/abs/1604.07316}
\showURL{%
\tempurl}


\bibitem[\protect\citeauthoryear{Catmull and Rom}{Catmull and Rom}{1974}]%
        {CatmullRom74}
\bibfield{author}{\bibinfo{person}{Edwin Catmull} {and}
  \bibinfo{person}{Raphael Rom}.} \bibinfo{year}{1974}\natexlab{}.
\newblock \showarticletitle{A Class of Local Interpolating Splines}.
\newblock In \bibinfo{booktitle}{\emph{Computer Aided Geometric Design}},
  \bibfield{editor}{\bibinfo{person}{R.~E. Barnhill} {and}
  \bibinfo{person}{R.~F. Riesenfeld}} (Eds.). \bibinfo{publisher}{Academic
  Press}, \bibinfo{pages}{317 -- 326}.
\newblock
\showISBNx{978-0-12-079050-0}
\urldef\tempurl%
\url{https://doi.org/10.1016/B978-0-12-079050-0.50020-5}
\showDOI{\tempurl}


\bibitem[\protect\citeauthoryear{Chen, Seff, Kornhauser, and Xiao}{Chen
  et~al\mbox{.}}{2015}]%
        {ChenSKC15}
\bibfield{author}{\bibinfo{person}{Chenyi Chen}, \bibinfo{person}{Ari Seff},
  \bibinfo{person}{Alain Kornhauser}, {and} \bibinfo{person}{Jianxiong Xiao}.}
  \bibinfo{year}{2015}\natexlab{}.
\newblock \showarticletitle{Deepdriving: Learning affordance for direct
  perception in autonomous driving}. In \bibinfo{booktitle}{\emph{Proceedings
  of the IEEE International Conference on Computer Vision}}.
  \bibinfo{pages}{2722--2730}.
\newblock


\bibitem[\protect\citeauthoryear{Clopper and Pearson}{Clopper and
  Pearson}{1934}]%
        {ClopperBiom34}
\bibfield{author}{\bibinfo{person}{C.~J. Clopper} {and} \bibinfo{person}{E.~S.
  Pearson}.} \bibinfo{year}{1934}\natexlab{}.
\newblock \showarticletitle{{The Use of Confidence or Fiducial Limits
  Illustrated in the Case of the Binomial}}.
\newblock \bibinfo{journal}{\emph{Biometrika}} \bibinfo{volume}{26},
  \bibinfo{number}{4} (\bibinfo{date}{12} \bibinfo{year}{1934}),
  \bibinfo{pages}{404--413}.
\newblock
\showISSN{0006-3444}
\urldef\tempurl%
\url{https://doi.org/10.1093/biomet/26.4.404}
\showDOI{\tempurl}


\bibitem[\protect\citeauthoryear{de~Jong}{de~Jong}{2004}]%
        {DeJong04}
\bibfield{author}{\bibinfo{person}{Edwin~D. de Jong}.}
  \bibinfo{year}{2004}\natexlab{}.
\newblock \showarticletitle{The Incremental Pareto-Coevolution Archive}. In
  \bibinfo{booktitle}{\emph{Genetic and Evolutionary Computation -- GECCO
  2004}}, \bibfield{editor}{\bibinfo{person}{Kalyanmoy Deb}} (Ed.).
  \bibinfo{publisher}{Springer Berlin Heidelberg}, \bibinfo{address}{Berlin,
  Heidelberg}, \bibinfo{pages}{525--536}.
\newblock
\showISBNx{978-3-540-24854-5}


\bibitem[\protect\citeauthoryear{{Deb}, {Pratap}, {Agarwal}, and
  {Meyarivan}}{{Deb} et~al\mbox{.}}{2002}]%
        {DebAM02}
\bibfield{author}{\bibinfo{person}{K. {Deb}}, \bibinfo{person}{A. {Pratap}},
  \bibinfo{person}{S. {Agarwal}}, {and} \bibinfo{person}{T. {Meyarivan}}.}
  \bibinfo{year}{2002}\natexlab{}.
\newblock \showarticletitle{A fast and elitist multiobjective genetic
  algorithm: NSGA-II}.
\newblock \bibinfo{journal}{\emph{IEEE Transactions on Evolutionary
  Computation}} \bibinfo{volume}{6}, \bibinfo{number}{2} (\bibinfo{date}{April}
  \bibinfo{year}{2002}), \bibinfo{pages}{182--197}.
\newblock
\urldef\tempurl%
\url{https://doi.org/10.1109/4235.996017}
\showDOI{\tempurl}


\bibitem[\protect\citeauthoryear{Fisher}{Fisher}{2006}]%
        {Fisher06}
\bibfield{author}{\bibinfo{person}{Ronald~Aylmer Fisher}.}
  \bibinfo{year}{2006}\natexlab{}.
\newblock \bibinfo{booktitle}{\emph{Statistical methods for research workers}}.
\newblock \bibinfo{publisher}{Genesis Publishing Pvt Ltd}.
\newblock


\bibitem[\protect\citeauthoryear{for Standardization~(ISO)}{for
  Standardization~(ISO)}{2019}]%
        {pas21448}
\bibfield{author}{\bibinfo{person}{International~Organization for
  Standardization~(ISO)}.} \bibinfo{year}{2019}\natexlab{}.
\newblock \bibinfo{title}{ISO/PAS 21448: Road vehicles -- Safety of the
  intended functionality}.
\newblock
\newblock


\bibitem[\protect\citeauthoryear{Fortin, De~Rainville, Gardner, Parizeau, and
  Gagn{\'e}}{Fortin et~al\mbox{.}}{2012}]%
        {FortinDGPG12}
\bibfield{author}{\bibinfo{person}{F{\'e}lix-Antoine Fortin},
  \bibinfo{person}{Fran\c{c}ois-Michel De~Rainville},
  \bibinfo{person}{Marc-Andr{\'e}~Gardner Gardner}, \bibinfo{person}{Marc
  Parizeau}, {and} \bibinfo{person}{Christian Gagn{\'e}}.}
  \bibinfo{year}{2012}\natexlab{}.
\newblock \showarticletitle{DEAP: Evolutionary Algorithms Made Easy}.
\newblock \bibinfo{journal}{\emph{J. Mach. Learn. Res.}} \bibinfo{volume}{13},
  \bibinfo{number}{1} (\bibinfo{date}{July} \bibinfo{year}{2012}),
  \bibinfo{pages}{2171--2175}.
\newblock
\showISSN{1532-4435}
\urldef\tempurl%
\url{http://dl.acm.org/citation.cfm?id=2503308.2503311}
\showURL{%
\tempurl}


\bibitem[\protect\citeauthoryear{Gambi, M{\"{u}}ller, and Fraser}{Gambi
  et~al\mbox{.}}{2019}]%
        {GambiMF19}
\bibfield{author}{\bibinfo{person}{Alessio Gambi}, \bibinfo{person}{Marc
  M{\"{u}}ller}, {and} \bibinfo{person}{Gordon Fraser}.}
  \bibinfo{year}{2019}\natexlab{}.
\newblock \showarticletitle{Automatically testing self-driving cars with
  search-based procedural content generation}. In
  \bibinfo{booktitle}{\emph{Proceedings of the 28th {ACM} {SIGSOFT}
  International Symposium on Software Testing and Analysis, {ISSTA}}}.
  \bibinfo{pages}{318--328}.
\newblock


\bibitem[\protect\citeauthoryear{Goodfellow, Bengio, and Courville}{Goodfellow
  et~al\mbox{.}}{2016}]%
        {GoodfellowMIT16}
\bibfield{author}{\bibinfo{person}{Ian~J. Goodfellow}, \bibinfo{person}{Yoshua
  Bengio}, {and} \bibinfo{person}{Aaron Courville}.}
  \bibinfo{year}{2016}\natexlab{}.
\newblock \bibinfo{booktitle}{\emph{Deep Learning}}.
\newblock \bibinfo{publisher}{MIT Press}, \bibinfo{address}{Cambridge, MA,
  USA}.
\newblock
\newblock
\shownote{\url{http://www.deeplearningbook.org}.}


\bibitem[\protect\citeauthoryear{Guo, Jiang, Zhao, Chen, and Sun}{Guo
  et~al\mbox{.}}{2018}]%
        {GuoJZCS18}
\bibfield{author}{\bibinfo{person}{Jianmin Guo}, \bibinfo{person}{Yu Jiang},
  \bibinfo{person}{Yue Zhao}, \bibinfo{person}{Quan Chen}, {and}
  \bibinfo{person}{Jiaguang Sun}.} \bibinfo{year}{2018}\natexlab{}.
\newblock \showarticletitle{DLFuzz: differential fuzzing testing of deep
  learning systems}. In \bibinfo{booktitle}{\emph{Proceedings of the 2018 {ACM}
  Joint Meeting on European Software Engineering Conference and Symposium on
  the Foundations of Software Engineering, {ESEC/SIGSOFT} {FSE}}}.
  \bibinfo{pages}{739--743}.
\newblock


\bibitem[\protect\citeauthoryear{Harman, Mansouri, and Zhang}{Harman
  et~al\mbox{.}}{2012}]%
        {HarmanMAZ12}
\bibfield{author}{\bibinfo{person}{Mark Harman}, \bibinfo{person}{S.~Afshin
  Mansouri}, {and} \bibinfo{person}{Yuanyuan Zhang}.}
  \bibinfo{year}{2012}\natexlab{}.
\newblock \showarticletitle{Search-based Software Engineering: Trends,
  Techniques and Applications}.
\newblock \bibinfo{journal}{\emph{ACM Comput. Surv.}} \bibinfo{volume}{45},
  \bibinfo{number}{1}, Article \bibinfo{articleno}{11} (\bibinfo{date}{Dec.}
  \bibinfo{year}{2012}), \bibinfo{numpages}{61}~pages.
\newblock
\showISSN{0360-0300}
\urldef\tempurl%
\url{https://doi.org/10.1145/2379776.2379787}
\showDOI{\tempurl}


\bibitem[\protect\citeauthoryear{Hazewinkel}{Hazewinkel}{1997}]%
        {Hazewinkel97}
\bibfield{author}{\bibinfo{person}{M. Hazewinkel}.}
  \bibinfo{year}{1997}\natexlab{}.
\newblock \bibinfo{booktitle}{\emph{Encyclopaedia of Mathematics: Supplement}}.
\newblock Number v. 1 in \bibinfo{series}{Encyclopaedia of Mathematics}.
  \bibinfo{publisher}{Springer Netherlands}.
\newblock
\showISBNx{9780792347095}
\showLCCN{87026437}


\bibitem[\protect\citeauthoryear{Heer and Bostock}{Heer and Bostock}{2010}]%
        {HeerB10}
\bibfield{author}{\bibinfo{person}{Jeffrey Heer} {and} \bibinfo{person}{Michael
  Bostock}.} \bibinfo{year}{2010}\natexlab{}.
\newblock \showarticletitle{Crowdsourcing Graphical Perception: Using
  Mechanical Turk to Assess Visualization Design}. In
  \bibinfo{booktitle}{\emph{Proceedings of the SIGCHI Conference on Human
  Factors in Computing Systems}} (Atlanta, Georgia, USA)
  \emph{(\bibinfo{series}{CHI '10})}. \bibinfo{publisher}{ACM},
  \bibinfo{address}{New York, NY, USA}, \bibinfo{pages}{203--212}.
\newblock
\showISBNx{978-1-60558-929-9}
\urldef\tempurl%
\url{https://doi.org/10.1145/1753326.1753357}
\showDOI{\tempurl}


\bibitem[\protect\citeauthoryear{Humbatova, Jahangirova, Bavota, Riccio,
  Stocco, and Tonella}{Humbatova et~al\mbox{.}}{2020}]%
        {HumbatovaICSE20}
\bibfield{author}{\bibinfo{person}{Nargiz Humbatova}, \bibinfo{person}{Gunel
  Jahangirova}, \bibinfo{person}{Gabriele Bavota}, \bibinfo{person}{Vincenzo
  Riccio}, \bibinfo{person}{Andrea Stocco}, {and} \bibinfo{person}{Paolo
  Tonella}.} \bibinfo{year}{2020}\natexlab{}.
\newblock \showarticletitle{Taxonomy of Real Faults in Deep Learning Systems}.
  In \bibinfo{booktitle}{\emph{Proceedings of 42nd International Conference on
  Software Engineering}} \emph{(\bibinfo{series}{ICSE ’20})}.
  \bibinfo{publisher}{ACM}, \bibinfo{pages}{12 pages}.
\newblock


\bibitem[\protect\citeauthoryear{ISO}{ISO}{2011}]%
        {iso26262}
\bibfield{author}{\bibinfo{person}{ISO}.} \bibinfo{year}{{2011}}\natexlab{}.
\newblock \bibinfo{title}{{Road vehicles -- Functional safety}}.
\newblock
\newblock


\bibitem[\protect\citeauthoryear{Kim, Feldt, and Yoo}{Kim
  et~al\mbox{.}}{2019}]%
        {KimFY19}
\bibfield{author}{\bibinfo{person}{Jinhan Kim}, \bibinfo{person}{Robert Feldt},
  {and} \bibinfo{person}{Shin Yoo}.} \bibinfo{year}{2019}\natexlab{}.
\newblock \showarticletitle{Guiding deep learning system testing using surprise
  adequacy}. In \bibinfo{booktitle}{\emph{Proceedings of the 41st International
  Conference on Software Engineering, {ICSE}}}. \bibinfo{pages}{1039--1049}.
\newblock


\bibitem[\protect\citeauthoryear{Kittur, Chi, and Suh}{Kittur
  et~al\mbox{.}}{2008}]%
        {KitturCS08}
\bibfield{author}{\bibinfo{person}{Aniket Kittur}, \bibinfo{person}{Ed~H. Chi},
  {and} \bibinfo{person}{Bongwon Suh}.} \bibinfo{year}{2008}\natexlab{}.
\newblock \showarticletitle{Crowdsourcing User Studies with Mechanical Turk}.
  In \bibinfo{booktitle}{\emph{Proceedings of the SIGCHI Conference on Human
  Factors in Computing Systems}} (Florence, Italy) \emph{(\bibinfo{series}{CHI
  '08})}. \bibinfo{publisher}{ACM}, \bibinfo{address}{New York, NY, USA},
  \bibinfo{pages}{453--456}.
\newblock
\showISBNx{978-1-60558-011-1}
\urldef\tempurl%
\url{https://doi.org/10.1145/1357054.1357127}
\showDOI{\tempurl}


\bibitem[\protect\citeauthoryear{Lakhotia, Harman, and McMinn}{Lakhotia
  et~al\mbox{.}}{2007}]%
        {LakhotiaHM07}
\bibfield{author}{\bibinfo{person}{Kiran Lakhotia}, \bibinfo{person}{Mark
  Harman}, {and} \bibinfo{person}{Phil McMinn}.}
  \bibinfo{year}{2007}\natexlab{}.
\newblock \showarticletitle{A Multi-objective Approach to Search-based Test
  Data Generation}. In \bibinfo{booktitle}{\emph{Proceedings of the 9th Annual
  Conference on Genetic and Evolutionary Computation}} (London, England)
  \emph{(\bibinfo{series}{GECCO '07})}. \bibinfo{publisher}{ACM},
  \bibinfo{address}{New York, NY, USA}, \bibinfo{pages}{1098--1105}.
\newblock
\showISBNx{978-1-59593-697-4}
\urldef\tempurl%
\url{https://doi.org/10.1145/1276958.1277175}
\showDOI{\tempurl}


\bibitem[\protect\citeauthoryear{Larman}{Larman}{1997}]%
        {Larman1997}
\bibfield{author}{\bibinfo{person}{Craig Larman}.}
  \bibinfo{year}{1997}\natexlab{}.
\newblock \bibinfo{booktitle}{\emph{Applying {UML} and Patterns: An
  Introduction to Object-Oriented Analysis and Design}}.
\newblock \bibinfo{publisher}{Prentice Hall}.
\newblock
\showISBNx{0-13-748880-7}


\bibitem[\protect\citeauthoryear{LeCun, Bottou, Bengio, Haffner,
  et~al\mbox{.}}{LeCun et~al\mbox{.}}{1998}]%
        {LecunBBH98}
\bibfield{author}{\bibinfo{person}{Yann LeCun}, \bibinfo{person}{L{\'e}on
  Bottou}, \bibinfo{person}{Yoshua Bengio}, \bibinfo{person}{Patrick Haffner},
  {et~al\mbox{.}}} \bibinfo{year}{1998}\natexlab{}.
\newblock \showarticletitle{Gradient-based learning applied to document
  recognition}.
\newblock \bibinfo{journal}{\emph{Proc. IEEE}} \bibinfo{volume}{86},
  \bibinfo{number}{11} (\bibinfo{year}{1998}), \bibinfo{pages}{2278--2324}.
\newblock


\bibitem[\protect\citeauthoryear{Lehman and Stanley}{Lehman and
  Stanley}{2011a}]%
        {LehmanS11}
\bibfield{author}{\bibinfo{person}{Joel Lehman} {and}
  \bibinfo{person}{Kenneth~O. Stanley}.} \bibinfo{year}{2011}\natexlab{a}.
\newblock \showarticletitle{Abandoning Objectives: Evolution Through the Search
  for Novelty Alone}.
\newblock \bibinfo{journal}{\emph{Evolutionary Computation}}
  \bibinfo{volume}{19}, \bibinfo{number}{2} (\bibinfo{year}{2011}),
  \bibinfo{pages}{189--223}.
\newblock
\urldef\tempurl%
\url{https://doi.org/10.1162/EVCO\_a\_00025}
\showDOI{\tempurl}


\bibitem[\protect\citeauthoryear{Lehman and Stanley}{Lehman and
  Stanley}{2011b}]%
        {LehmanS11b}
\bibfield{author}{\bibinfo{person}{Joel Lehman} {and}
  \bibinfo{person}{Kenneth~O. Stanley}.} \bibinfo{year}{2011}\natexlab{b}.
\newblock \showarticletitle{Evolving a Diversity of Virtual Creatures Through
  Novelty Search and Local Competition}. In
  \bibinfo{booktitle}{\emph{Proceedings of the 13th Annual Conference on
  Genetic and Evolutionary Computation}} (Dublin, Ireland)
  \emph{(\bibinfo{series}{GECCO '11})}. \bibinfo{publisher}{ACM},
  \bibinfo{address}{New York, NY, USA}, \bibinfo{pages}{211--218}.
\newblock
\showISBNx{978-1-4503-0557-0}
\urldef\tempurl%
\url{https://doi.org/10.1145/2001576.2001606}
\showDOI{\tempurl}


\bibitem[\protect\citeauthoryear{Levenshtein}{Levenshtein}{1966}]%
        {LevenshteinSov66}
\bibfield{author}{\bibinfo{person}{Vladimir~I Levenshtein}.}
  \bibinfo{year}{1966}\natexlab{}.
\newblock \showarticletitle{Binary codes capable of correcting deletions,
  insertions, and reversals}. In \bibinfo{booktitle}{\emph{Soviet physics
  doklady}}, Vol.~\bibinfo{volume}{10}. \bibinfo{pages}{707--710}.
\newblock


\bibitem[\protect\citeauthoryear{Ma, Juefei-Xu, Zhang, Sun, Xue, Li, Chen, Su,
  Li, Liu, Zhao, and Wang}{Ma et~al\mbox{.}}{2018}]%
        {Ma-ASE-2018}
\bibfield{author}{\bibinfo{person}{Lei Ma}, \bibinfo{person}{Felix Juefei-Xu},
  \bibinfo{person}{Fuyuan Zhang}, \bibinfo{person}{Jiyuan Sun},
  \bibinfo{person}{Minhui Xue}, \bibinfo{person}{Bo Li},
  \bibinfo{person}{Chunyang Chen}, \bibinfo{person}{Ting Su},
  \bibinfo{person}{Li Li}, \bibinfo{person}{Yang Liu}, \bibinfo{person}{Jianjun
  Zhao}, {and} \bibinfo{person}{Yadong Wang}.} \bibinfo{year}{2018}\natexlab{}.
\newblock \showarticletitle{DeepGauge: Multi-granularity Testing Criteria for
  Deep Learning Systems}. In \bibinfo{booktitle}{\emph{Proceedings of the 33rd
  ACM/IEEE International Conference on Automated Software Engineering}}
  (Montpellier, France) \emph{(\bibinfo{series}{ASE 2018})}.
  \bibinfo{publisher}{ACM}, \bibinfo{address}{New York, NY, USA},
  \bibinfo{pages}{120--131}.
\newblock
\showISBNx{978-1-4503-5937-5}
\urldef\tempurl%
\url{https://doi.org/10.1145/3238147.3238202}
\showDOI{\tempurl}


\bibitem[\protect\citeauthoryear{Mao, Capra, Harman, and Jia}{Mao
  et~al\mbox{.}}{2017}]%
        {MaoCHJ17}
\bibfield{author}{\bibinfo{person}{Ke Mao}, \bibinfo{person}{Licia Capra},
  \bibinfo{person}{Mark Harman}, {and} \bibinfo{person}{Yue Jia}.}
  \bibinfo{year}{2017}\natexlab{}.
\newblock \showarticletitle{A survey of the use of crowdsourcing in software
  engineering}.
\newblock \bibinfo{journal}{\emph{Journal of Systems and Software}}
  \bibinfo{volume}{126} (\bibinfo{year}{2017}), \bibinfo{pages}{57--84}.
\newblock
\urldef\tempurl%
\url{https://doi.org/10.1016/j.jss.2016.09.015}
\showDOI{\tempurl}


\bibitem[\protect\citeauthoryear{Mao, Harman, and Jia}{Mao
  et~al\mbox{.}}{2016}]%
        {MaoHJ16}
\bibfield{author}{\bibinfo{person}{Ke Mao}, \bibinfo{person}{Mark Harman},
  {and} \bibinfo{person}{Yue Jia}.} \bibinfo{year}{2016}\natexlab{}.
\newblock \showarticletitle{Sapienz: Multi-objective Automated Testing for
  Android Applications}. In \bibinfo{booktitle}{\emph{Proceedings of the 25th
  International Symposium on Software Testing and Analysis}}
  (Saarbr\&\#252;cken, Germany) \emph{(\bibinfo{series}{ISSTA 2016})}.
  \bibinfo{publisher}{ACM}, \bibinfo{address}{New York, NY, USA},
  \bibinfo{pages}{94--105}.
\newblock
\showISBNx{978-1-4503-4390-9}
\urldef\tempurl%
\url{https://doi.org/10.1145/2931037.2931054}
\showDOI{\tempurl}


\bibitem[\protect\citeauthoryear{{Marculescu}, {Feldt}, and
  {Torkar}}{{Marculescu} et~al\mbox{.}}{2016}]%
        {MarculescuFT2016}
\bibfield{author}{\bibinfo{person}{B. {Marculescu}}, \bibinfo{person}{R.
  {Feldt}}, {and} \bibinfo{person}{R. {Torkar}}.}
  \bibinfo{year}{2016}\natexlab{}.
\newblock \showarticletitle{Using Exploration Focused Techniques to Augment
  Search-Based Software Testing: An Experimental Evaluation}. In
  \bibinfo{booktitle}{\emph{2016 IEEE International Conference on Software
  Testing, Verification and Validation (ICST)}}. \bibinfo{pages}{69--79}.
\newblock
\urldef\tempurl%
\url{https://doi.org/10.1109/ICST.2016.26}
\showDOI{\tempurl}


\bibitem[\protect\citeauthoryear{Mouret}{Mouret}{2011}]%
        {Mouret09}
\bibfield{author}{\bibinfo{person}{Jean-Baptiste Mouret}.}
  \bibinfo{year}{2011}\natexlab{}.
\newblock \showarticletitle{Novelty-Based Multiobjectivization}. In
  \bibinfo{booktitle}{\emph{New Horizons in Evolutionary Robotics}},
  \bibfield{editor}{\bibinfo{person}{St{\'e}phane Doncieux},
  \bibinfo{person}{Nicolas Bred{\`e}che}, {and} \bibinfo{person}{Jean-Baptiste
  Mouret}} (Eds.). \bibinfo{publisher}{Springer Berlin Heidelberg},
  \bibinfo{address}{Berlin, Heidelberg}, \bibinfo{pages}{139--154}.
\newblock
\showISBNx{978-3-642-18272-3}


\bibitem[\protect\citeauthoryear{Mouret and Clune}{Mouret and Clune}{2015}]%
        {Mouret15}
\bibfield{author}{\bibinfo{person}{Jean-Baptiste Mouret} {and}
  \bibinfo{person}{Jeff Clune}.} \bibinfo{year}{2015}\natexlab{}.
\newblock \bibinfo{title}{Illuminating search spaces by mapping elites}.
\newblock
\newblock
\showeprint[arxiv]{cs.AI/1504.04909}


\bibitem[\protect\citeauthoryear{Nelder and Wedderburn}{Nelder and
  Wedderburn}{1972}]%
        {NelderStat72}
\bibfield{author}{\bibinfo{person}{J.~A. Nelder} {and}
  \bibinfo{person}{R.~W.~M. Wedderburn}.} \bibinfo{year}{1972}\natexlab{}.
\newblock \showarticletitle{Generalized Linear Models}.
\newblock \bibinfo{journal}{\emph{Journal of the Royal Statistical Society:
  Series A (General)}} \bibinfo{volume}{135}, \bibinfo{number}{3}
  (\bibinfo{year}{1972}), \bibinfo{pages}{370--384}.
\newblock
\urldef\tempurl%
\url{https://doi.org/10.2307/2344614}
\showDOI{\tempurl}


\bibitem[\protect\citeauthoryear{of~State~Highway and
  Officials}{of~State~Highway and Officials}{2018}]%
        {aashto2018}
\bibfield{author}{\bibinfo{person}{American~Association of State~Highway} {and}
  \bibinfo{person}{Transportation Officials}.} \bibinfo{year}{2018}\natexlab{}.
\newblock \bibinfo{booktitle}{\emph{AASHTO Green Book (GDHS-7) - A Policy on
  Geometric Design of Highways and Streets}}.
\newblock \bibinfo{publisher}{American Association of State Highway and
  Transportation Officials}.
\newblock
\showISBNx{9781560516767}


\bibitem[\protect\citeauthoryear{Panichella, Kifetew, and Tonella}{Panichella
  et~al\mbox{.}}{2018}]%
        {PanichellaKT18}
\bibfield{author}{\bibinfo{person}{Annibale Panichella},
  \bibinfo{person}{Fitsum~Meshesha Kifetew}, {and} \bibinfo{person}{Paolo
  Tonella}.} \bibinfo{year}{2018}\natexlab{}.
\newblock \showarticletitle{Automated Test Case Generation as a Many-Objective
  Optimisation Problem with Dynamic Selection of the Targets}.
\newblock \bibinfo{journal}{\emph{{IEEE} Transactions on Software Engineering}}
  \bibinfo{volume}{44}, \bibinfo{number}{2} (\bibinfo{year}{2018}),
  \bibinfo{pages}{122--158}.
\newblock


\bibitem[\protect\citeauthoryear{{Pastore}, {Mariani}, and {Fraser}}{{Pastore}
  et~al\mbox{.}}{2013}]%
        {PastoreMF13}
\bibfield{author}{\bibinfo{person}{F. {Pastore}}, \bibinfo{person}{L.
  {Mariani}}, {and} \bibinfo{person}{G. {Fraser}}.}
  \bibinfo{year}{2013}\natexlab{}.
\newblock \showarticletitle{CrowdOracles: Can the Crowd Solve the Oracle
  Problem?}. In \bibinfo{booktitle}{\emph{2013 IEEE Sixth International
  Conference on Software Testing, Verification and Validation}}.
  \bibinfo{pages}{342--351}.
\newblock
\urldef\tempurl%
\url{https://doi.org/10.1109/ICST.2013.13}
\showDOI{\tempurl}


\bibitem[\protect\citeauthoryear{Peer, Vosgerau, and Acquisti}{Peer
  et~al\mbox{.}}{2014}]%
        {PeerVA14}
\bibfield{author}{\bibinfo{person}{Eyal Peer}, \bibinfo{person}{Joachim
  Vosgerau}, {and} \bibinfo{person}{Alessandro Acquisti}.}
  \bibinfo{year}{2014}\natexlab{}.
\newblock \showarticletitle{Reputation as a sufficient condition for data
  quality on Amazon Mechanical Turk}.
\newblock \bibinfo{journal}{\emph{Behavior Research Methods}}
  \bibinfo{volume}{46}, \bibinfo{number}{4} (\bibinfo{date}{01 Dec}
  \bibinfo{year}{2014}), \bibinfo{pages}{1023--1031}.
\newblock
\showISSN{1554-3528}
\urldef\tempurl%
\url{https://doi.org/10.3758/s13428-013-0434-y}
\showDOI{\tempurl}


\bibitem[\protect\citeauthoryear{Pei, Cao, Yang, and Jana}{Pei
  et~al\mbox{.}}{2017}]%
        {PeiCYJ17}
\bibfield{author}{\bibinfo{person}{Kexin Pei}, \bibinfo{person}{Yinzhi Cao},
  \bibinfo{person}{Junfeng Yang}, {and} \bibinfo{person}{Suman Jana}.}
  \bibinfo{year}{2017}\natexlab{}.
\newblock \showarticletitle{DeepXplore: Automated Whitebox Testing of Deep
  Learning Systems}. In \bibinfo{booktitle}{\emph{Proceedings of the 26th
  Symposium on Operating Systems Principles}}. \bibinfo{pages}{1--18}.
\newblock


\bibitem[\protect\citeauthoryear{Selinger}{Selinger}{2003}]%
        {Selinger03}
\bibfield{author}{\bibinfo{person}{P. Selinger}.}
  \bibinfo{year}{2003}\natexlab{}.
\newblock \showarticletitle{Potrace: a polygon-based tracing algorithm}.
\newblock  (\bibinfo{year}{2003}).
\newblock
\urldef\tempurl%
\url{http://potrace.sourceforge.net/potrace.pdf}
\showURL{%
\tempurl}


\bibitem[\protect\citeauthoryear{Tian, Pei, Jana, and Ray}{Tian
  et~al\mbox{.}}{2018}]%
        {TianPSB18}
\bibfield{author}{\bibinfo{person}{Yuchi Tian}, \bibinfo{person}{Kexin Pei},
  \bibinfo{person}{Suman Jana}, {and} \bibinfo{person}{Baishakhi Ray}.}
  \bibinfo{year}{2018}\natexlab{}.
\newblock \showarticletitle{DeepTest: Automated Testing of
  Deep-neural-network-driven Autonomous Cars}. In
  \bibinfo{booktitle}{\emph{Proceedings of the 40th International Conference on
  Software Engineering}} (Gothenburg, Sweden) \emph{(\bibinfo{series}{ICSE
  '18})}. \bibinfo{publisher}{ACM}, \bibinfo{address}{New York, NY, USA},
  \bibinfo{pages}{303--314}.
\newblock
\showISBNx{978-1-4503-5638-1}
\urldef\tempurl%
\url{https://doi.org/10.1145/3180155.3180220}
\showDOI{\tempurl}


\bibitem[\protect\citeauthoryear{\tool}{\tool}{2019}]%
        {tool}
\tool \bibinfo{year}{2019}\natexlab{}.
\newblock \bibinfo{title}{\tool: A Tool for Model-based Exploration of the
  Frontier of Behaviours for Deep Learning Systems Testing}.
\newblock
  \bibinfo{howpublished}{\url{https://github.com/testingautomated-usi/DeepJanus}}.
\newblock


\bibitem[\protect\citeauthoryear{{Unity Technologies}}{{Unity
  Technologies}}{2019}]%
        {unitygameengine}
\bibfield{author}{\bibinfo{person}{{Unity Technologies}}.}
  \bibinfo{year}{2019}\natexlab{}.
\newblock \bibinfo{booktitle}{\emph{Unity}}.
\newblock
\urldef\tempurl%
\url{https://unity.com}
\showURL{%
\tempurl}


\bibitem[\protect\citeauthoryear{Wicker, Huang, and Kwiatkowska}{Wicker
  et~al\mbox{.}}{2018}]%
        {WickerHK18}
\bibfield{author}{\bibinfo{person}{Matthew Wicker}, \bibinfo{person}{Xiaowei
  Huang}, {and} \bibinfo{person}{Marta Kwiatkowska}.}
  \bibinfo{year}{2018}\natexlab{}.
\newblock \showarticletitle{Feature-Guided Black-Box Safety Testing of Deep
  Neural Networks}. In \bibinfo{booktitle}{\emph{Tools and Algorithms for the
  Construction and Analysis of Systems - 24th International Conference,
  {TACAS}}}. \bibinfo{pages}{408--426}.
\newblock


\bibitem[\protect\citeauthoryear{Yoo and Harman}{Yoo and Harman}{2007}]%
        {YooH07}
\bibfield{author}{\bibinfo{person}{Shin Yoo} {and} \bibinfo{person}{Mark
  Harman}.} \bibinfo{year}{2007}\natexlab{}.
\newblock \showarticletitle{Pareto Efficient Multi-objective Test Case
  Selection}. In \bibinfo{booktitle}{\emph{Proceedings of the 2007
  International Symposium on Software Testing and Analysis}} (London, United
  Kingdom) \emph{(\bibinfo{series}{ISSTA '07})}. \bibinfo{publisher}{ACM},
  \bibinfo{address}{New York, NY, USA}, \bibinfo{pages}{140--150}.
\newblock
\showISBNx{978-1-59593-734-6}
\urldef\tempurl%
\url{https://doi.org/10.1145/1273463.1273483}
\showDOI{\tempurl}


\bibitem[\protect\citeauthoryear{Yoo and Harman}{Yoo and Harman}{2010}]%
        {YooH10}
\bibfield{author}{\bibinfo{person}{Shin Yoo} {and} \bibinfo{person}{Mark
  Harman}.} \bibinfo{year}{2010}\natexlab{}.
\newblock \showarticletitle{Using hybrid algorithm for Pareto efficient
  multi-objective test suite minimisation}.
\newblock \bibinfo{journal}{\emph{Journal of Systems and Software}}
  \bibinfo{volume}{83}, \bibinfo{number}{4} (\bibinfo{year}{2010}),
  \bibinfo{pages}{689 -- 701}.
\newblock
\showISSN{0164-1212}
\urldef\tempurl%
\url{https://doi.org/10.1016/j.jss.2009.11.706}
\showDOI{\tempurl}


\bibitem[\protect\citeauthoryear{Zhang, Zhang, Zhang, Liu, and Khurshid}{Zhang
  et~al\mbox{.}}{2018}]%
        {ZhangZZ0K18}
\bibfield{author}{\bibinfo{person}{Mengshi Zhang}, \bibinfo{person}{Yuqun
  Zhang}, \bibinfo{person}{Lingming Zhang}, \bibinfo{person}{Cong Liu}, {and}
  \bibinfo{person}{Sarfraz Khurshid}.} \bibinfo{year}{2018}\natexlab{}.
\newblock \showarticletitle{DeepRoad: GAN-based metamorphic testing and input
  validation framework for autonomous driving systems}. In
  \bibinfo{booktitle}{\emph{Proceedings of the 33rd {ACM/IEEE} International
  Conference on Automated Software Engineering, {ASE}}}.
  \bibinfo{pages}{132--142}.
\newblock


\end{thebibliography}

\end{document}